\documentclass[11pt,a4paper,oneside]{amsart}
\usepackage[foot]{amsaddr}

\usepackage[
hmarginratio={1:1},     
vmarginratio={1:1},     
textwidth=450pt,        
textheight=700pt,
heightrounded,          
]{geometry}
\usepackage{graphicx}
\usepackage{amsmath,amsfonts}
\usepackage{natbib}
\usepackage{booktabs}

\usepackage{bm}
\usepackage{bbm}

\RequirePackage[colorlinks,citecolor=blue,urlcolor=blue]{hyperref}
\usepackage{etoolbox}

\makeatletter
\patchcmd{\@makecaption}
  {\parbox}
  {\advance\@tempdima-\fontdimen2} 
  {}{}
\makeatother

\title[Estimating seal pup production]{Estimating seal pup production in the Greenland Sea using Bayesian hierarchical modeling}
\author{Martin Jullum\textsuperscript{1}}
\address{\textsuperscript{1,2}Norwegian Computing Center, Oslo, Norway.}
\email{jullum@nr.no}
\author{Thordis Thorarinsdottir\textsuperscript{2}}
\email{thordis@nr.no}
\author[Jullum, Thorarinsdottir \& Bachl]{Fabian E. Bachl\textsuperscript{3}}
\address{\textsuperscript{3}School of Mathematics, University of Edinburgh,
Edinburgh,
Scotland.}
\email{bachlfab@gmail.com}

\newcommand{\hatt}[1]{\widehat{#1}}

\newcommand{\dd}{\, \mathrm{d}}

\newcommand{\infl}[3]{{\rm{IF}}_{#1}(#2;#3)}

\newcommand{\empinfl}[2]{\infl{\mu}{Y_i}{\Gnhat}}

\newcommand{\Gnhat}{\widehat{G}_n}

\newcommand{\N}{{\rm{N}}}

\newcommand{\mockalph}[1]{}

\begin{document}

\begin{abstract}
The Greenland Sea is an important breeding ground for harp and hooded seals. Estimates of the annual seal pup production are critical factors in the abundance estimation needed for management of the species. These estimates are usually based on counts from aerial photographic surveys. However, only a minor part of the whelping region can be photographed, due to its large extent. To estimate the total seal pup production, we propose a Bayesian hierarchical modeling approach motivated by viewing the seal pup appearances as a realization of a log-Gaussian Cox process using covariate information from satellite imagery as a proxy for ice thickness. For inference, we utilize the stochastic partial differential equation (SPDE) module of the integrated nested  Laplace approximation (INLA) framework. In a case study using survey data from 2012, we compare our results with existing methodology in a comprehensive cross-validation study. The results of the  study indicate that our method improves local estimation performance, and that the increased prediction uncertainty of our method is required to obtain calibrated count predictions. This suggests that the sampling density of the survey design may not be sufficient to obtain reliable estimates of the seal pup production.
\end{abstract}

\maketitle

\section{Introduction}
\label{sec:Intro}

Three stocks of harp seals (Pagophilus groenlandicus) and two (possibly three) stocks of hooded seals (Cystophora cristata) inhabit the North Atlantic Ocean where they have been harvested for centuries \citep{Sergeant1974, Sergeant1991, KovacsLavigne1986}. Monitoring the abundance of seals is vital for controlling the biodiversity in the region. State-of-the-art seal population models are dynamically built based on historical catch data \citep{oigaard2014current, oigaard2014pup}. The main ingredient in these models is the total pup production in a given year which needs to be quantified based on on-site observational data since other quantification methods based on catch-at-age and mark-recapture data etc.~are considered unreliable \citep{ICES2014}. The whelping regions in the North Atlantic typically cover several thousand square kilometers so that the total pup production needs to be estimated based on observations from a minor part of the region. Both the estimated total pup production \textit{and} the associated uncertainty are then used as input in dynamic population models \citep{oigaard2010estimation}. 

The observational data consists of seal pup counts obtained by manual counting on photographs. The photographs stem from an aerial photographic survey conducted by flying along transects sparsely covering the whelping region. The survey methodology is discussed in more detail in Section \ref{sec:Data}.  The traditional method for estimating the total pup production based on such count data is that of \citet{kingsley1985distribution} which assumes a homogeneous dispersion of seals across the entire whelping region.  \citet{salberg2009estimation} propose a generalized additive modeling (GAM) approach \citep{hastie1990generalized}, assuming the counts follow a negative binomial distribution and taking the spatial location of the counts into account. For data that is close to homogeneous, the negative binomial GAM approach and the Kingsley method yield similar estimates. However, the Kingsley method may possess a positive bias when the spatial distribution of the pups is clustered \citep{salberg2008estimation, oigaard2010estimation}. Additionally, the GAM method produces much smaller uncertainty bounds than the homogeneous Kingsley approach. 

In this paper, we propose a new method for estimating the total seal pup production. We view the seal pup appearances as a spatial point process \citep{moller2003statistical} and model the point pattern of the seal pups as a log-Gaussian Cox process \citep[LGCP;][]{moller1998log}  with a spatial latent field which also allows additional covariate information to be accounted for.
In a Bayesian formulation with priors on the model parameters, the seal pup production estimate is represented by the posterior predictive distribution found by integrating the posterior distribution over the spatial domain of the whelping region, instead of a single point estimate accompanied with a variance estimate. 
This Bayesian hierarchical model can be fitted by utilizing the stochastic partial differential equation (SPDE) approach of the integrated nested Laplace approximation method \citep[INLA;][]{lindgren2011explicit,Rue2009}. The final posterior predictive distribution can subsequently be computed from this fitted model by a sampling approach. Although more traditional Markov Chain Monte Carlo (MCMC) methods in theory could be used to arrive at the same posterior, application of INLA allows results to be produced magnitudes faster,
at a negligible cost in terms of accuracy.

To illustrate and test this methodology, we use seal pup photo counts from an aerial photographic survey in the Greenland Sea in March 2012 with two different types of seals, harp and hooded seals. The data set contains  the spatial location of each photo and the corresponding pup count. To be more informative about the non-observed areas, we include covariate information extracted from satellite imagery captured on the very same date as the aerial photographic survey was conducted, to act as a proxy for ice thickness. This is important as the seal pups can only be observed on ice, with non-observable pups accounted for within the dynamic population model
\citep{oigaard2010estimation}.
Compared to the other procedures our method gives larger uncertainties, especially for the harp seals. To validate these differences, we compare our proposed method with a number of reference methods in two cross-validation experiments, one where random sets of photos are removed and one where whole transects are removed from the data set prior to inference. Performance assessment based on proper scoring rules suggests our method performs best on a local level, and comparable on a more global scale. Further calibration assessment suggests that the larger uncertainty in our method is indeed more realistic.

The rest of the paper is organized as follows: Section \ref{sec:Data} describes the survey method used to gather seal pup observational data, in addition to specific details related to our particular seal pup data set. The satellite imagery and the covariate information extracted therefrom are also discussed. Relevant background related to point processes and aggregated point patterns are given in Section \ref{sec:Background}. Section \ref{sec:Methods} describes the details of our suggested modeling approach and three references methods that we compare our method against. The validation schemes used to verify and compare the different approaches are also described. The results are presented in Section \ref{sec:Results}, including specifics of the model fitted with our procedure, and the validation and comparison results. The final Section \ref{sec:ConcludingRemarks} contains concluding remarks and pointers to future work.

\section{Data}
\label{sec:Data}

In this section we describe the survey method and additional modeling information we have obtained through satellite imagery. 

\subsection{Survey method}
\label{sec:surveying method}

Before conducting the aerial photographic survey with the purpose of monitoring the seal pup production, the marine researchers typically perform a helicopter reconnaissance survey. This is done in order to locate the patches where the seals whelp for limiting the survey area for the more expensive airborne photographic survey. The actual photographic survey is conducted by flying a survey aircraft equipped with advanced photographic equipment and GPS along a number of transects at a fixed distance that sparsely cover the survey area. 

In this particular survey in March 2012, the airplane flew at an altitude of about 330m, and took a total of 2792 photos along 27 parallel transects, approximately 3Nm ($\approx$ 5.6km) apart, with each photo covering $226 \times 346$m of ground level. Due to fog, an exception was made for the two southernmost transects, which were flown at an altitude of 250m, with the photos covering $170 \times 260$m. Care is taken to record non-overlapping photos. Some overlaps appear in the data, but since the overlaps are so small and completely irrelevant for practical purposes, we assume non-overlap in the subsequent analysis. Along each transect, the cameras were turned on when the first seal was spotted from the airplane, and photos were taken continuously until the ice edge was reached on the eastern side, and until no seals were spotted for an extended period to the west. As a consequence of this survey setup, the whelping region is approximately defined as the union of the 1.5Nm ($\approx 2.8$km) bands around each transect. Thus, when estimating the total pup production in the whelping region, we only count predictions within this area. More details about the survey may be found in \citet{oigaard2014current,oigaard2014pup}. 

\subsection{Seal pup counts}
\label{sec:sealpupcount}

\begin{figure}[ht!]
	\centering
	\makebox{\includegraphics[width=0.60\columnwidth]{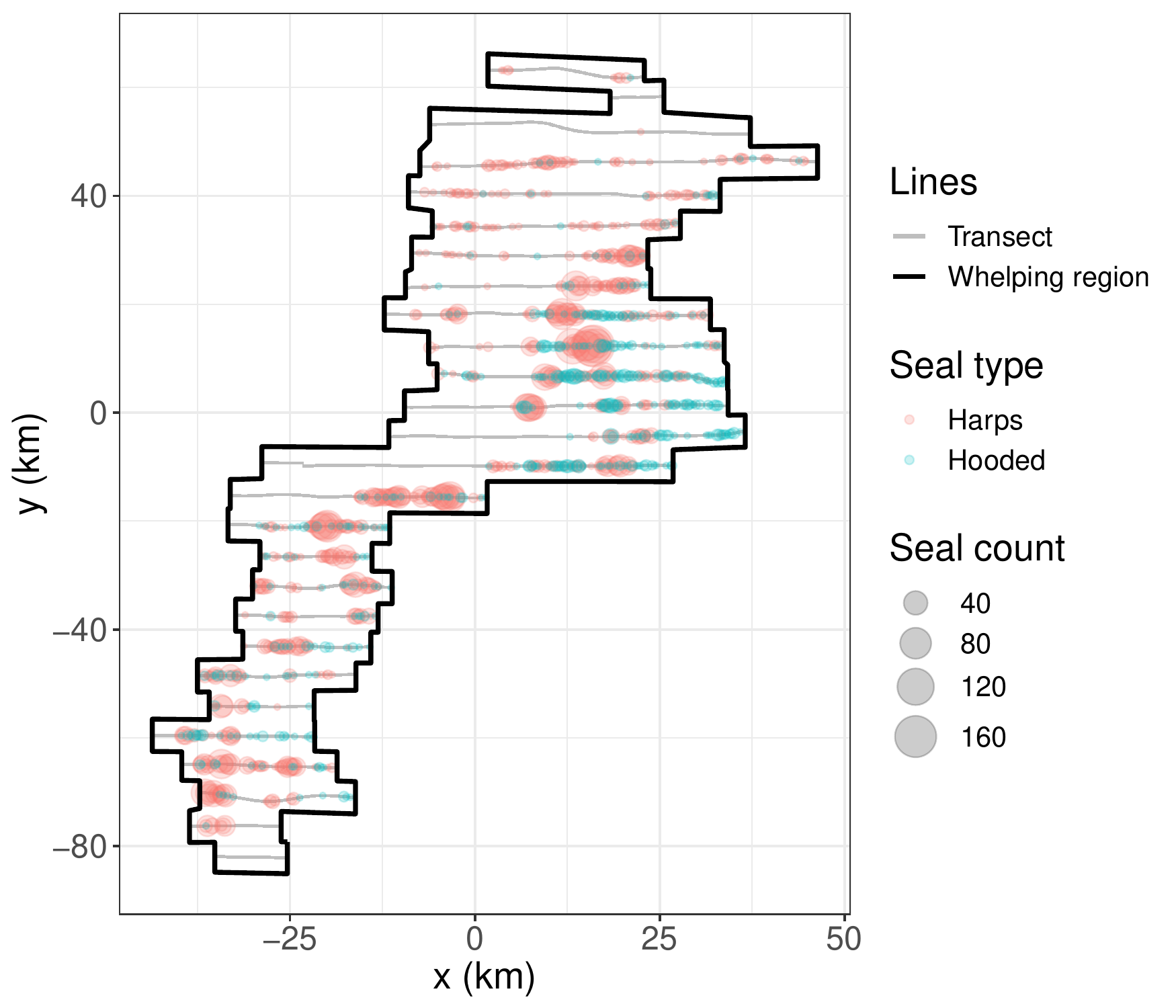}}
	\caption{Harp and hooded seal pup count data from the Greenland Sea 2012 survey, with east-west direction on the $x$-axis and north-south direction on the $y$-axis.  }
	\label{fig:sealdata}
\end{figure}

Following the airborne survey, experienced marine researchers manually count the number of seal pups of each species in each photo\footnote{The exact position of the seals are not recorded.}. Quality checks with multiple examinations are performed to limit the measurement error introduced in this step \citep{oigaard2014current,oigaard2014pup}. The seal pup count data set used on the subsequent analysis contains the coordinates and extent of each photo, in addition to the number of seal pups of each species observed. The data are plotted in Figure~\ref{fig:sealdata} along with the transect locations and the extent of the whelping region. As seen from the figure, there tends to be more seal pups clustered towards the middle eastern boundary and southern corner of the whelping region. Comparatively more harp seal pups are observed than hooded seal pups and the spatial distribution of the former seems less homogeneous. 

\subsection{Satellite imagery}
\label{sec:satellite imagery}

\begin{figure}[ht!]
	\centering
	\makebox{\includegraphics[width=0.60\columnwidth]{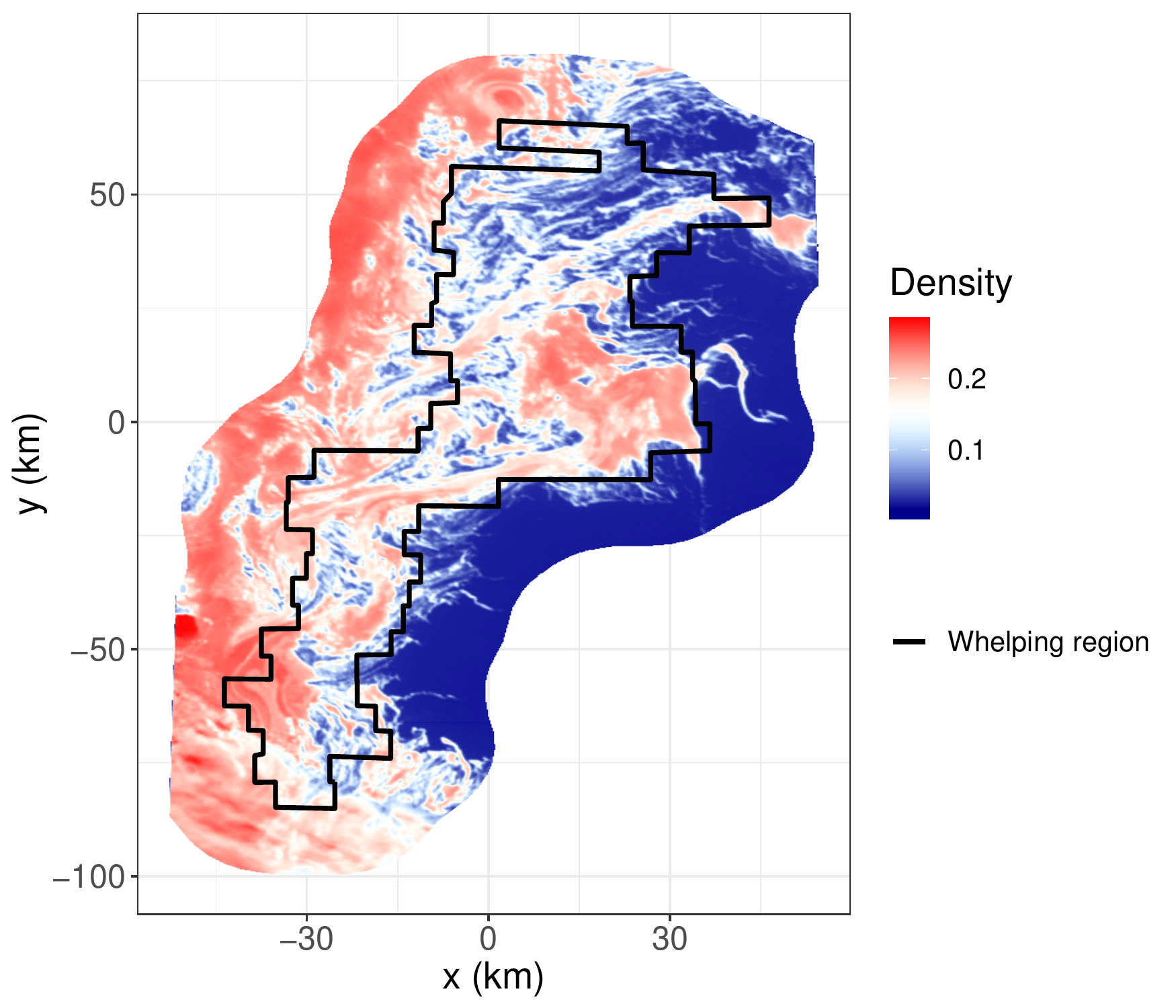}}
	\caption{Satellite data showing ice density used as a covariate in the model fitting process.}
	\label{fig:satellitedata}
\end{figure}

For whelping, the seals require large ice floes with access to the ocean for the adult seals to access food. Information regarding which areas are covered by ice floes and which areas are merely open water, is thus potentially highly relevant when estimating the seal pup production. In an attempt to account for this, we have collected high resolution satellite imagery (Modis) from the whelping region captured on the same day as the airborne photographic survey was conducted. From this satellite imagery we have extracted a variable which acts as a proxy for the ice thickness. This density variable is displayed in Figure~\ref{fig:satellitedata}. Comparing the satellite data to the seal pup counts in Figure~\ref{fig:sealdata}, we see that the seal pup counts appear to be higher in the areas with high ice density, than in areas with lower ice density.

\section{Background}
\label{sec:Background}

In the present paper, we assume that the seal pup appearances can be thought of as a realization of a point process. The aim of the analysis is to estimate the total number of seal pups, $N(\Omega)$, in the whelping region $\Omega \subset \mathbb{R}^2$ (indicated in black in Figures~\ref{fig:sealdata} and \ref{fig:satellitedata}).

\subsection{Point processes}

A spatial stochastic point process is a mathematical description of the random process through which points or locations are distributed in space. The collection of such observations is called a point pattern. In informal mathematical terms, a spatial point process $Y$ is a random collection of points in a bounded observation region $\Omega \subset \mathbb{R}^2$ where both the number of points and their locations are random. The most fundamental type of point process is the \textit{Poisson point process}. It may be specified by a deterministic intensity function $\lambda: \Omega \mapsto [0,\infty)$ which determines the density of points in any location in $\Omega$. The number of points, $N(B)$, of any Borel set $B \subseteq \Omega$ is Poisson distributed with mean $\mu(B) = \int_B \lambda(\bm{s})\dd \bm{s}$, i.e. $N(B) \sim \text{Po}(\mu(B))$. Further, $N(B)$ is independent of $N(B^*)$ for any other non-overlapping Borel set $B^* \subseteq \Omega$ with $N(B \cup B^*) = N(B) + N(B^*)$. 

The {\em Cox process} (also known as the doubly stochastic Poisson point process) introduced by \citet{cox1955some} is a generalization of the Poisson point process where the intensity function $\lambda$ is in itself stochastic. A popular special case of this hierarchical model is the {\em log-Gaussian Cox process} (LGCP) \citep{moller1998log}, where the intensity is assumed to be log-Gaussian; i.e.~there is an underlying latent Gaussian random field $Z$, and given $\lambda=\exp(Z)$, $Y$ is a Poisson point process with intensity $\lambda$. This type of model is quite flexible and useful for modeling a great variety of natural processes, in particular when only a single realization of the process is available \citep[see e.g.][]{diggle2013statistical, illian2008statistical, moller2003statistical}. The LGCP model can describe various types of clustering in the point pattern through the positive semi-definite correlation function of the underlying Gaussian process \citep{moller1998log}.  

\citet{WolpertIckstadt98} consider a Cox process where the stochastic intensity function is given by $\lambda = \Gamma$ for a latent gamma random field $\Gamma$. As before, the conditional distribution of $N(B)$ given $\lambda$ is a Poisson distribution. For this model, the marginal distribution of $N(B)$ is available in closed form and given by a negative binomial distribution \citep{Matern1971, DiggleMilne1983}. 

\subsection{Models for aggregated point patterns}

As mentioned in Section \ref{sec:sealpupcount}, the exact positions of the counted seal pups are not available in the survey data. Instead, the seal pup counts are provided as aggregated counts per photo. Thus, our data is a partly observed point pattern aggregated to counts on an irregular lattice, as opposed to the actual point pattern. Methods that fit doubly stochastic Poisson process models to point pattern data, such as those proposed by \citet{WolpertIckstadt98}, \citet{Simpson&2016} and \citet{Yuan&2017}, are thus not directly applicable. Rather, our setting can be viewed as an extension of the approach considered in \citet{Rue2009} where a fully observed point pattern is approximated by the corresponding counts on a \textit{regular} lattice. 

To this end, let $N(A_i)$ for $i=1,\ldots,n$ denote the number of seal pups in each of the $n=2792$ photos with domains $A_i \subset \Omega$ and (potentially varying) areas $|A_i|$ for $i=1,\ldots,n$. Per properties of Poisson driven point processes, conditional on the intensity $\lambda$, being either deterministic (Poisson process) or random (Cox process), we have that $N(A_i)$ is Poisson distributed with parameter $\mu=\int_{A_i}\lambda(\bm{s})\dd \bm{s}$, i.e.
\begin{align}
N(A_i) \, | \, \lambda \sim \text{Po}\Big(\mu = \int_{A_i}\lambda(\bm{s})\dd \bm{s}\Big), \qquad i=1,\ldots,n. \label{eq:NiZ}
\end{align}
Working under a Bayesian paradigm, we obtain an estimate of the total number of seal pups in $\Omega$ given by the posterior predictive distribution of $N(\Omega)$,  
\begin{align}
\label{eq:postpreddist}
\begin{split}
p\big(N(\Omega) \, | \, N(A_1),\ldots,N(A_n)\big) &= \int p\big(N(\Omega), \lambda \, | \, N(A_1),\ldots,N(A_n)\big) \dd \lambda  \\
&= \int p\big(N(\Omega) \, | \, \lambda\big)p\big( \lambda \, | \, N(A_1),\ldots,N(A_n)\big) \dd \lambda, 
\end{split}
\end{align}
where $N(\Omega) \, | \, \lambda \sim \text{Po}(\mu = \int_{\Omega} \lambda(s)\dd \bm{s})$.
The estimate for $N(\Omega)$ is thus given by a mixture of Poisson distributions rather than a single Poisson distribution under both deterministic and the random models for the intensity function $\lambda$. As the data are overdispersive, this may improve the fit of the Poisson model. 

The negative binomial distribution is a common choice for modeling random counts in applications with clustering. Due to the overdispersion in the data, \citet{salberg2009estimation} model the seal pup counts using a generalized additive model (GAM) based on a negative binomial likelihood. \cite{DiggleMilne1983} show that only two point process models yield a negative binomial marginal count distribution. One such model is the so-called compound Poisson process where each point of an underlying Poisson process is replaced by a random number of coincident points where the numbers are independent and identically distributed according to a logarithmic distribution. The second example is the Cox process with an intensity function given by a gamma random field considered by \citet{WolpertIckstadt98}. Similarly, if a Poisson process has a constant intensity and the inference is performed using a conjugate gamma prior distribution for $\lambda$, the resulting posterior predictive distribution in \eqref{eq:postpreddist} is a negative binomial distribution, see Section~\ref{sec:HomPo} below.     

\section{Methods}
\label{sec:Methods}

We consider four different approaches to modeling the seal pup counts, see the summary in Table~\ref{tab:models}. Our new proposed method is based on the LGCP framework and accounts for potential clustering, or overdispersion, in the data through the underlying latent Gaussian random field. This approach is compared to the following reference approaches: A GAM approach previously proposed by \citet{salberg2009estimation} and \citet{oigaard2010estimation} under both a Poisson and a negative binomial likelihood, and a homogeneous Poisson model. The methods and the associated inference approaches are described below.  


\begin{table}\caption{Summary of fitted models and inference approaches. Here, $\bm{s}$ is a spatial index, $\lambda_0, \alpha$ and $\bm{\beta}$ are coefficients, $\bm{x}$ is a vector of covariates, $f$ denotes a latent spatial random effect, and $S$ a deterministic spatial smoothing component.\label{tab:models}} 
	\makebox{\begin{tabular}{llll}
			\toprule
			Name     & Model & Intensity & Estimation \\ \midrule
			LGCP & Poisson  & $\exp(\alpha + \bm{\beta}^\top \bm{x}(\bm{s}) + f(\bm{s}))$     & SPDE-INLA  \\
			GAM Po & Poisson  & $\exp(\alpha + \bm{\beta}^\top \bm{x}(\bm{s}) + S(\bm{s}))$     & GAM  \\
			GAM NB & Negative Binomial  & $\exp(\alpha + \bm{\beta}^\top \bm{x}(\bm{s}) + S(\bm{s}))$     & GAM  \\
			Hom Po & Poisson  & $\lambda_0$ & Bayesian   \\ \bottomrule
	\end{tabular}}
\end{table}

\subsection{Stochastic intensity model}
\label{sec:stoch.int.model}

Our proposed approach (LGCP) is based on a Poisson point process with a stochastic intensity function (i.e.~a Cox process). The stochastic intensity function takes the form
\begin{equation}\label{eq:stoch intensity}
\lambda(\bm{s}) = \exp(Z_f(\bm{s})),
\end{equation}
where $Z_f$ is a Gaussian random field, or an LGCP with a conditional distribution for the aggregated counts on the form of \eqref{eq:NiZ}. The continuous Gaussian random field takes the form
\begin{align}
Z_f(\bm{s}) = \alpha + \bm{\beta}^\top \bm{x}(\bm{s}) + f(\bm{s}), \label{eq:Zf}
\end{align}
where $\alpha$ is an intercept term, $\bm{\beta}$ are regression coefficients for $\bm{x}(\bm{s}) = (q(\bm{s}),s_1,s_2,\sqrt{s_1^2 + s_2^2})^\top$, with $q(\bm{s})$ containing the ice density variable from the satellite imagery, while $s_1, s_2,$ and $\sqrt{s_1^2 + s_2^2}$ model linear spatial effects for $\bm{s} = (s_1, s_2)^\top \in \Omega$. Finally, $f(\bm{s})$ is a (non-linear) continuous Gaussian random field meant to model spatial dependence not captured by the covariates in $\bm{x}$. Specifically, $f(\bm{s})$ is given the Mat\`{e}rn covariance function of the form
\begin{align}
\text{Cov}(f(\bm{s}),f(\bm{t})) = \frac{\sigma^2}{2^{\nu-1}\Gamma(\nu)}(\kappa\|\bm{s}-\bm{t}\|)^\nu K_{\nu}(\kappa\|\bm{s}-\bm{t}\|), \label{eq:Matern}
\end{align}
for $\bm{s}, \bm{t} \in \Omega$ where $\nu>0$ is a smoothing parameter, $K_{\nu}$ is the modified Bessel function of the second kind, $\kappa>0$ is a scaling parameter and $\sigma^2$ is the marginal variance. Further, for identifiability of the intercept term in \eqref{eq:Zf}, we restrict $f(\bm{s})$ to integrate to zero over the modeling region.

Note that the model defined by \eqref{eq:NiZ}, \eqref{eq:stoch intensity} and \eqref{eq:Zf} is equivalent to the Poisson log-normal model of \citet{ChristensenWaagepetersen2002} which is a special case of the spatial generalized linear mixed model framework proposed by \citet{Diggle&1998}. \citet{ChristensenWaagepetersen2002} employ a Markov chain Monte Carlo (MCMC) algorithm for model inference using a slightly simpler covariance structure than the Mat\'ern covariance function in \eqref{eq:Matern} to analyze a data set where the observation sets $\{A_i\}_{i=1}^n$ are given by discs of fixed radius. Similar methods based on Poisson kriging are discussed in e.g. \citet{Bellier&2010} and \citet{DeOliveira2014} where more traditional geostatistical inference methods are employed. 

\subsubsection{Inference}

To obtain an approximation to the posterior predictive distribution in \eqref{eq:postpreddist}, we apply the integrated nested Laplace approximation (INLA) of \citet{Rue2009} that allows for computationally feasible approximate Bayesian inference with \textit{discrete} space latent Gaussian models, see Appendix \ref{app:INLA} for a brief description. The stochastic partial differential equation (SPDE) approach of \citet{lindgren2011explicit} extends the INLA framework to also handle models with \textit{continuous} latent fields as in \eqref{eq:Zf}. The SPDE approach is based on transforming the continuous latent field to a certain Gaussian Markov random field (GMRF), formulated through the solution of a SPDE. The key point is to approximate the continuous field $Z(\bm{s})$ by a field $Z_{\text{GMRF}}(\bm{s})$ living on a triangular mesh. For a triangular mesh with $m$ triangle vertices, we write 
\begin{align}
Z_{\text{GMRF}}(\bm{s}) = \sum_{j=1}^m z_j \phi_j(\bm{s}), \label{eq:Z.FEM}
\end{align}
where $\bm{z}=(z_1,\ldots,z_m)^\top$ is a multivariate Gaussian random vector and $\{\phi_j(\bm{s})\}^m_{j=1}$ is a set of deterministic linearly independent basis functions which are piecewise linear between the vertices and chosen such that $\phi_j(\bm{s})$ is 1 at vertex $j$, and 0 at all other vertices. A consequence of the representation in \eqref{eq:Z.FEM} is that $Z_{\text{GMRF}}(\bm{s})$ is fully determined by $\bm{z}$. That is, $Z_{\text{GMRF}}(\bm{s})$ takes the value $z_j$ at vertex $j$ while its values inside the triangles are determined by linear interpolation.

Assume now that $Z(\bm{s})$ is equipped with the Mat\`{e}rn covariance function in \eqref{eq:Matern}. As this type of field is a solution to a certain SPDE, the precision matrix $Q$ of $\bm{z}$ takes an analytical form which can be approximated by a sparse matrix $\tilde{Q}$. Since $Z_{\text{GMRF}}(\bm{s})$ is completely determined by $\bm{z}$, this allows continuous field computations to be carried out approximately using the INLA implementation. Note that following certain guidelines for constructing the triangular mesh, the resulting approximation error is typically small \citep{lindgren2011explicit,simpson2012order}. 
For a complete introduction and review of the INLA framework, including the SPDE approach, see e.g. \citet{RINLAbook2015} and \citet{rue2016bayesian}. 

\subsubsection{Model specification and fitting}

The model parameters to be estimated are the regression parameters $(\alpha, \bm{\beta})$ in \eqref{eq:Zf} and the hyperparameters $(\nu, \kappa, \sigma^2)$ of the Mat\'ern covariance function in \eqref{eq:Matern}. To improve identifiability, we fix the Mat\`{e}rn smoothing parameter at $\nu=2$, as is common when applying the INLA framework \citep[e.g.][]{RINLAbook2015}. The other hyperparameters related to the latent field are equipped with mesh dependent default priors specified by the INLA software. For our mesh these are $\theta_1 \sim \mathcal{N}(1.328,10)$ and $\theta_2 \sim \mathcal{N}(-2.594,10)$, where $\theta_1 = \log(\tau)$ and $\theta_2 = \log(\kappa)$, with $\sigma^2 = 1/(4\pi\kappa^2\tau^2)$.  The intercept term $\alpha$ is assigned the improper prior $\mathcal{N}(0,\infty)$ while $\bm{\beta} \sim \mathcal{N}(0,1000 \, \mathcal{I}_4)$, where $\mathcal{I}_4$ denotes the $4\times 4$ dimensional identity matrix. All these priors are non-informative and they seem to influence the final results to a very limited degree. The alternative use of penalized complexity (PC) priors \citep{simpson2017penalising} resulted in very similar final results.

\begin{figure}
	\centering
	\makebox{\includegraphics[width=0.60\columnwidth]{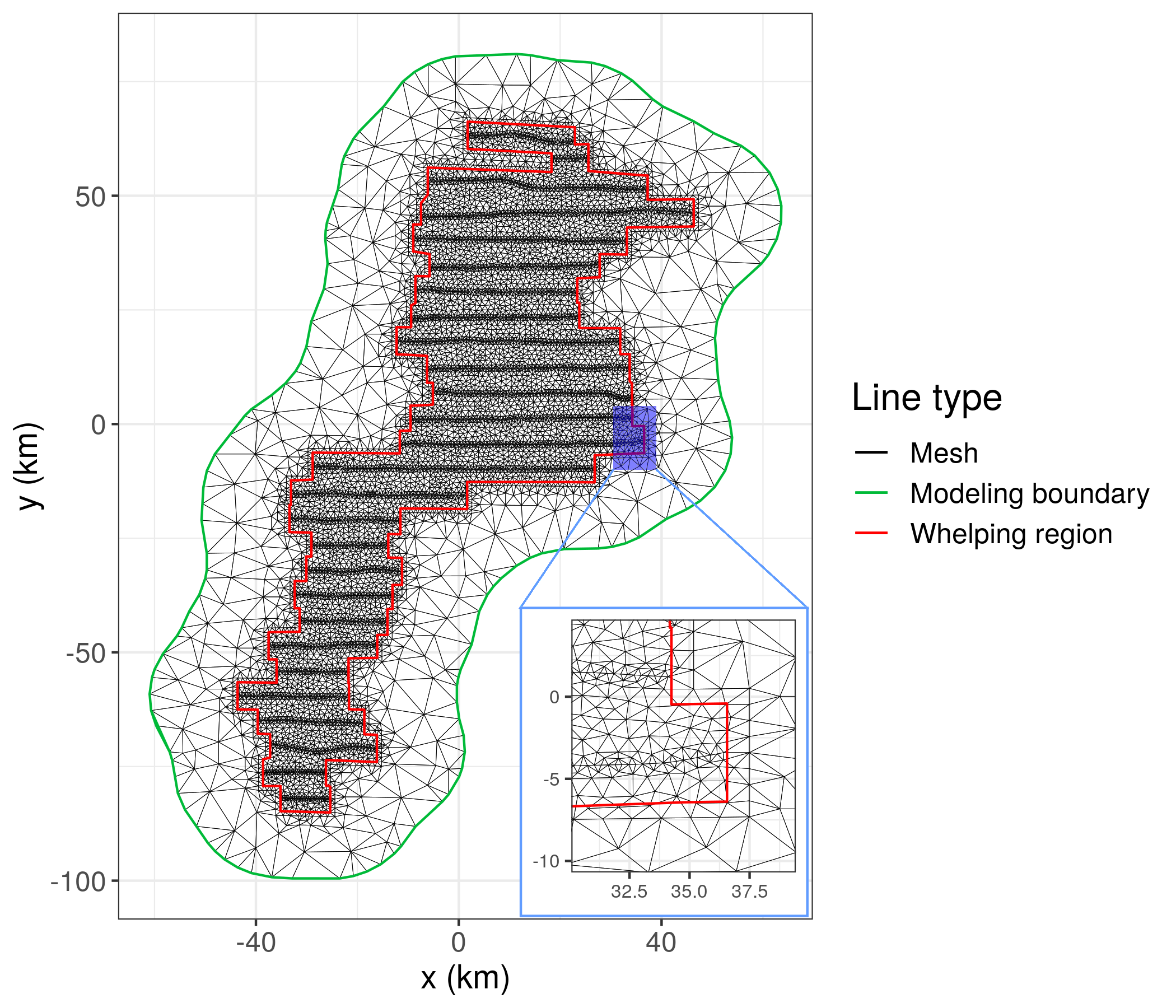}} 
	\caption{The triangular mesh used in the SPDE-INLA analysis for the LGCP model. The bottom right corner shows a zoomed in version of the mesh for the blue area above.}
	\label{fig:mesh}
\end{figure}

The triangular mesh we use to fit the LGCP model with SPDE-INLA is displayed in Figure \ref{fig:mesh}. To overcome boundary effects, we extend the area that is modeled quite a bit beyond the whelping region as recommended by \citet{lindgren2011explicit}. In order to properly represent observations being aggregated over a certain spatial domain (the photos), we construct the mesh in a specific way such that there is a mesh node in the center of every photo and that the corresponding Voronoi tessellation \citep{watson1981computing} matches the photo. The Voronoi tessellation is used to specify the weight or ``offset'' of the observations used in the Poisson distribution, see e.g. \citet[Ch. 4.1]{krainski2018advanced} which refers to this as the ``dual mesh''. 
Voronoi tessellations that match the photos are obtained by placing mesh nodes at the center point of each photo, and at a distance equal to the height of the photo both above and below the center point. For respectively, the leftmost and rightmost photo on each transect, additional points are placed to the left and the right, at a distance equal to the width of each photo.
Note that this procedure is merely a technical task carried out in order to fit the problem with the SPDE approach while still using the exact locations of the irregular lattice (the photos) in a Poisson regression formulation. The mesh then has a coarser resolution elsewhere, where fine resolution detail of the latent field cannot be easily estimated. 

The INLA software produces posterior distributions for all individual hyperparameters, and enables sampling from the posterior of the complete latent field
\[
p( Z \, | \, N(A_1),\ldots,N(A_n)),
\]
see \citet[][Ch.~8.2]{RINLAbook2015}. This allows us to use a Monte Carlo approximation for the integral in \eqref{eq:postpreddist}:
\begin{align}
\int p(N(\Omega)| Z)p( Z | N(A_1),\ldots,N(A_n)) \dd Z &\approx \frac{1}{K}\sum_{k=1}^K p(N(\Omega)| \tilde{Z}_k), \notag
\end{align}
where $\tilde{Z}_k$ is the $k$-th sample of the posterior latent field. We use $K=10\,000$ in this and similar Monte Carlo integrations throughout the paper. Further, by the point process properties,
\[
p(N(\Omega) \, | \, Z = \tilde{Z}_k) \sim \text{Po}\Big(\mu = \int_\Omega \exp(\tilde{Z}_k(\bm{s})) \dd \bm{s}\Big).
\]
The integral $\int_\Omega \exp(\tilde{Z}_k(\bm{s})) \dd \bm{s}$ can be solved by e.g.~a simple Riemann midpoint rule. This is achieved by dividing $\Omega$ into $J \approx 40\,000$ rectangles $B_1,\ldots,B_J$ centered in $\bm{s}_1,\ldots,\bm{s}_J$, similar in size to the individual photos  $\{A_i\}_{i=1}^n$, and using $\int_\Omega \exp(\tilde{Z}_k(\bm{s})) \dd \bm{s} \approx \sum_{j=1}^J \exp(\tilde{Z}_k(\bm{s}_j))|B_j|$ where the values $\tilde{Z}_k(\bm{s}_j)$ are derived using \eqref{eq:Z.FEM}. To find the $J$ sets, the full modeling region is covered by a regular grid and the cells within the whelping region are selected. The final approximation to the posterior predictive distribution is thus
\begin{align}
p\big(N(\Omega)\, | \, N(A_1),\ldots,N(A_n)\big) &\approx \frac{1}{K} \sum_{k=1}^K \text{Po}\Big(\mu=\sum_{j=1}^J \exp(\tilde{Z}_k(\bm{s}_j))\, |B_j|\Big), \label{eq:postpreddist.approx}
\end{align}
i.e.~a Poisson mixture distribution.

\subsection{Inhomogeneous intensity model}
\label{sec:inhom.int.model}
An alternative to the stochastic intensity function in \eqref{eq:Zf} is a deterministic inhomogeneous intensity. With only one realization of the point pattern, it is generally not possible to distinguish between a doubly stochastic Poisson process and an inhomogenous Poisson process \citep{moller2003statistical}. However, the underlying assumptions regarding the data structure (the LGCP e.g.~assumes clustering) and the components of the intensity model vary somewhat so that the resulting predictive distributions may differ.  

The models we use here are motivated by the work of \citet{salberg2009estimation} who model the seal pup production by a generalized additive model (GAM) based on a negative binomial likelihood, except that they do not include covariate information. Apart from adding covariates, the below approach follows that of \citet{salberg2009estimation} and \citet{oigaard2010estimation}. \citet{salberg2009estimation} argue that a negative binomial likelihood should be used rather than a Poisson likelihood due to overdispersion in that data.  However, as the inhomogeneous intensity and the Poisson mixture in the posterior predictive distribution may improve the fit of the Poisson model, we find it natural to include the inhomogenous Poisson model in our list of models. 

\subsubsection{Poisson model}
\label{sec:Pois.inhomogen.model}
The inhomogeneous Poisson model takes the conditional Poisson form of \eqref{eq:NiZ} with intensity function $\lambda(\bm{s})=\exp(\eta(\bm{s}))$ given by 
\begin{align}
\eta(\bm{s}) = \alpha + \bm{\beta}^\top \bm{x}(\bm{s}) + S(\bm{s}), \label{eq:mu.GAM}
\end{align}
where $S(\cdot)$ is a spatial smoothing component given by a thin-plate smoothing regression spline \citep{wood2003thin-plate}, i.e.~a smooth, nonlinear deterministic spatial effect. As noted by \citet[][Ch. 5.8]{wood2017generalized} there is a certain duality between the smoothing component $S(\cdot)$ in \eqref{eq:mu.GAM} and the Gaussian random field $f(\cdot)$ in \eqref{eq:Zf} in that $S(\cdot)$ can be interpreted as the mean of a latent random field. An assessment of uncertainty in the estimation of $S(\cdot)$ then corresponds to an assessment of the uncertainty in the mean of the random field. 

To fit the model in \eqref{eq:mu.GAM}, we follow \citet{salberg2009estimation}, and rely on the \verb,gam, function in the {\tt R}-package \verb,mgcv, \citep{wood2017generalized}. This function fits the model
\[
N(A_i) \sim \text{Po}\big(\mu = |A_i|\exp(\eta(\bm{s}_i))\big), \quad \text{for } i=1,\ldots,n,
\]
with the $|A_i|$ values as fixed offsets with overlapping cubic regressions on a set of artificial knots in space, and $A_i$ centered in $\bm{s}_i$. A generalized cross-validation (GCV) criterion is used to select the right amount of smoothing.

The method essentially returns maximum likelihood (ML) estimates for the regression parameters $\alpha, \bm{\beta}$ and parameters $\bm{\nu}$ associated with the spatial smoothing component $S(\cdot)$, $\bm{\theta} = (\alpha, \bm{\beta}, \bm{\nu})$. To account for the uncertainty involved in the Poisson model we rely on the asymptotic distribution of $\bm{\theta}$,
\begin{align}
N\big(\bm{\theta},\text{Cov}(\hatt{\bm{\theta}})\big), \label{eq:Asymp.Norm}
\end{align}
where $\text{Cov}(\hatt{\bm{\theta}})$ is the asymptotic covariance of $\hatt{\bm{\theta}}$ estimated by \verb,gam,. We sample from  \eqref{eq:Asymp.Norm} and approximate the integral over $\Omega$ by a simple Riemann midpoint rule using the $J$ rectangles $B_1,\ldots,B_J$ centered in $\bm{s}_1,\ldots,\bm{s}_J$ with $B_1 \cup \ldots \cup B_J \approx \Omega$, as in \eqref{eq:postpreddist.approx}. This gives the approximation
\begin{align}
p_{\text{GAM Po}}\big(N(\Omega) \, | \, N(A_1),\ldots,N(A_n) \big) \approx \frac{1}{K}\sum_{k=1}^K \text{Po}\Big(\mu = \sum_{j=1}^J|B_j| \exp(\eta_{\tilde{\theta}_k}(\bm{s}_j))\Big), \label{eq:GAMPois.approx}
\end{align}
where $\eta_{\tilde{\theta}_k}(\bm{s}_j)$, is the value of \eqref{eq:mu.GAM} using $\bm{\theta}$-parameters corresponding to the $k$-th sample from \eqref{eq:Asymp.Norm}. 

\subsubsection{Negative binomial model}
\label{sec:NegBin.inhomogen.model}

An alternative to the inhomogeneous Poisson model is to consider the negative binomial model
\begin{align}
N(A_i)|\lambda \sim \text{NegBin}\Big(\mu = \int_{A_i} \lambda(\bm{s}) \dd \bm{s} ,\tau \Big), \label{eq:GAM.orig}
\end{align}
with $\lambda(\bm{s})=\exp(\eta(\bm{s}))$ where $\eta(\bm{s})$ is given in \eqref{eq:mu.GAM}, and shape parameter $\tau$. This model is fit in the same manner as the Poisson model above using
\[
N(A_i)|\lambda \sim \text{NegBin}\big(\mu = |A_i|\exp(\eta(\bm{s}_i)),\tau \big), \quad \text{for } i=1,\ldots,n.
\]

In contrast to the Poisson distribution, the negative binomial distribution is not closed under addition for different mean values. Thus, to arrive at a full posterior predictive distribution for the total seal pup count $N(\Omega)$ under the model in \eqref{eq:GAM.orig}, we employ a sampling procedure that relies on an underlying conditional independence condition stating that conditional on $\mu$ and $\tau$, the point counts in disjoint sets are independent. Again, using the $J$ disjoint sets $B_1,\ldots,B_J$, we sample counts 
\[
N_{k}(B_j)|\lambda \sim \text{NegBin}\Big(\mu = |B_j| \exp(\eta_{\tilde{\theta}_k}(\bm{s}_j)), \tau = \hatt{\tau}_j \Big), \quad \text{for } j=1,\ldots,J; k=1,\ldots,K,
\]
where $\bm{s}_j$ is the center point of $B_j$, $\eta_{\tilde{\theta}_k}(\bm{s}_j)$ is sampled from \eqref{eq:Asymp.Norm} as in \eqref{eq:GAMPois.approx}, and $\hatt{\tau}_j$ is the estimated shape parameter for $B_j$. The posterior predictive distribution for $N(\Omega)$ is then given by the empirical distribution function of $\{N_k(\Omega)\}_{k=1}^K$ where
\[
N_k(\Omega) = \sum_{j=1}^J N_k(B_j), \quad \text{for } k = 1,\ldots,K.
\]
Both the sampling from the asymptotic normal distribution in \eqref{eq:Asymp.Norm} and the conditional independence assumption employed above, are also used by \citet{salberg2009estimation} to quantify the uncertainty around the total seal pup production.

\subsection{Homogeneous intensity model}\label{sec:HomPo}

A simple reference model is a homogeneous Poisson model with a constant intensity. That is, we set $\lambda(\bm{s}) \equiv \lambda_{0}$ for a fixed scalar $\lambda_0$ and assume
\begin{align}
N(A_i)|\lambda \sim \text{Po}(\mu = |A_i|\lambda_0), \quad \text{for } i=1,\ldots,n. \notag
\end{align}
For a Bayesian inference, we equip $\lambda_0$ with a non-informative conjugate gamma prior, 
\[\lambda_0 \sim \Gamma(a_0 = 10,b_0 = 10),\]
for both seal types, where $a_0$ and $b_0$ are the shape and rate parameters, respectively.
This results in the following posterior distribution,
\begin{align}
\lambda_0|N(A_1),\ldots,N(A_n) \sim \Gamma\Big(a = a_0 + \sum_{i=1}^n N(A_i),
b = b_0 + \sum_{i=1}^n |A_i| \Big). \notag 
\end{align}
Moreover, the posterior predictive distribution in \eqref{eq:postpreddist} takes the form of a negative binomial distribution and is available in a closed form
\begin{align}
p_{\text{Hom Po}}\big(N(\Omega) \, | \, N(A_1),\ldots,N(A_n)\big) 
&=\int \text{Po}(|\Omega|\lambda_0) \Gamma(a ,b) \dd \lambda_0 \notag \\
&= \text{NegBin}(\mu = |\Omega|a/b, \tau = a).
\label{eq:HomPo.postpred}
\end{align}
where $a$ and $b$ are specified in the posterior distribution for $\lambda_0$ above. 

The motivation for the homogeneous Poisson model is not only that it is a simple special case of our proposed LGCP model, but also that it is similar to the traditional approach to estimate seal pup production based on aerial photographic transect surveys, often referred to as Kingsley's method \citep{kingsley1985distribution}. Kingsley's method is fundamentally simple: For each transect $T_1,\ldots,T_{27}$ covering the space $A_{T_k} = \bigcup_{\{A_i \subset T_k\}} A_i$, compute the seal pup count $N_{T_k} = \sum_{A_i \in T_k} N(A_i)$. Then, the estimate of the total number of seal pups is
\begin{align}
\hatt{N}(\Omega) = \frac{|A_{\Omega}|}{\sum_{k=1}^{27} |A_{T_k}|}\sum_{k=1}^{27} N_{T_k}. \label{eq:Kingsley}
\end{align}
\citet{kingsley1985distribution} also provide an estimate of the variance related to the seal pup production estimate, based on serial differences between the transects. \citet{salberg2008estimation} later provided a modification to this variance estimate, which we have used here. Since this method only provides a point and a variance estimate, it is difficult to properly compare it against the remaining Bayesian procedures. We will therefore not perform validation tests, as described in Section \ref{sec:verification}, for this method.

\subsection{Verification}
\label{sec:verification}

We compare the various modeling approaches using a cross-validation scheme where we rely on two procedures for subsetting the data.  The first procedure is a standard 10-fold cross-validation setup, where we randomly remove 10\% of the photos each time, such that each photo is removed exactly once. In the second procedure we remove all photos in one full transect at a time, such that each transect is removed exactly once, leaving us with 27 different subsets. For both procedures, we fit the competing models for every subset and compute posterior predictive distributions for every photo that is removed along with posterior predictive distributions for the sum of the removed photos (corresponding to the full transect for the latter procedure).

We compare the predictive performance of the various modeling approaches using two performance measures: The logarithmic score \citep{good1952rational} and the continuous ranked probability score (CRPS) \citep{matheson1976scoring}, which both are proper scoring rules that assess full predictive distributions \citep{gneiting2007strictly}. Denoting a generic posterior predictive distribution by $g(x)$, the corresponding cumulative distribution function by $G(x)$, and the observed count by $y_\text{true}$, the two performance measures takes the form
\begin{align}
\text{logScore}(g, y_\text{true}) & = -\log(g(y_{\text{true}})), \label{eq:logScore_CRPS} \\
\text{CRPS}(G, y_\text{true}) & = \int_{-\infty}^{\infty} (G(x) - \mathbbm{1}_{\{x > y_{\text{true}}\}}(x))^2 \dd x,  \notag
\end{align} 
where $\mathbbm{1}_{\{\cdot\}}(x)$ denotes the indicator function. For both measures, smaller values reflect a better model.

\begin{figure}
	\centering
	\makebox{\includegraphics[width=0.80\columnwidth]{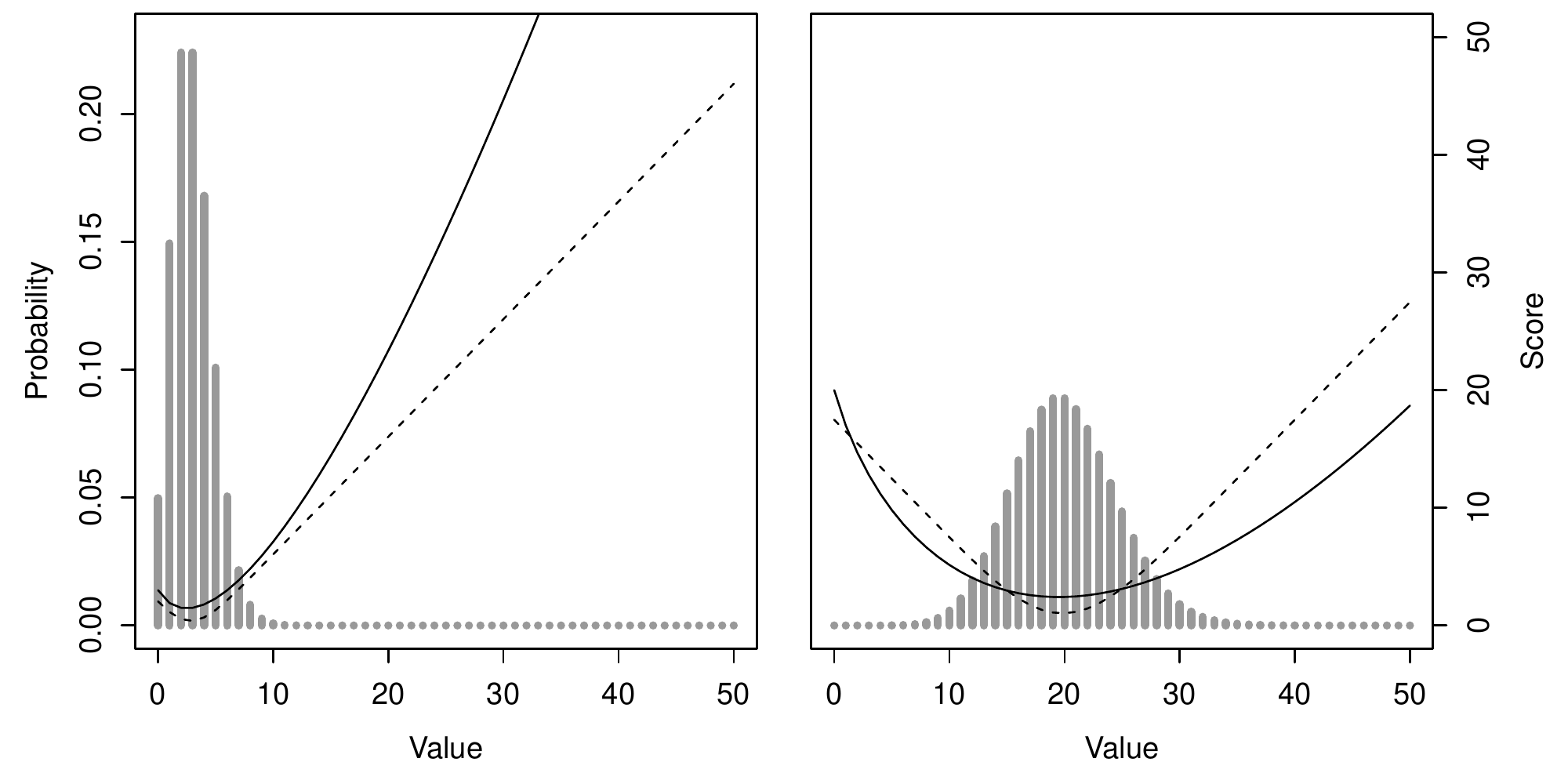}} 
	\caption{
		Illustration of the logScore (solid lines) and CRPS (dashed lines) functions for different values of $y_{\text{true}}$,
		with two different predictive distributions (indicated in gray), Po($\mu=3$) (left plot) and Po($\mu=20$) (right plot). 
	}
	\label{fig:scores}
\end{figure}

Both the logarithmic score and the CRPS are optimized in expectation when the true data distribution is issued as the forecast \citep{gneiting2007strictly}. However, in real-life situations, average scores are often associated with high uncertainty \citep{thorarinsdottir2019verification}. We thus apply two scores that penalize prediction errors in slightly different manners, see Figure~\ref{fig:scores}. In particular, the logarithmic score is more sensitive to outliers, cf.~the left plot in Figure~\ref{fig:scores}. For every cross-validation schemes, we present mean scores, where the mean is computed over the different folds. To indicate the variation in the scores, we compute 90\% bootstrapped confidence intervals (CI) of those means
by repeated re-sampling (10\,000 times, with replacement) of the fold scores.

We further assess the calibration, or the prediction uncertainty, of the methods by assessing the coverage of the posterior predictive distributions. That is, we check how often $y_{\text{true}}$ lies within different credibility intervals, compared to their intended coverage -- small uncertainty is of no value if it is not reflecting the true uncertainty of the model. For a calibrated forecast, an event predicted with probability $p$ should be realized with the same frequency in the observed data. In order to mimic quantification of the prediction uncertainty for the complete whelping region as closely as possible, we perform this exercise on the transect level. 

\section{Results}
\label{sec:Results}

Here, we present the results obtained when applying the various models/estimation methods in Table \ref{tab:models} to per-photo count data from the 2012 survey of the Greenland sea whelping region. We model hooded and harp seals separately as their occurrences are expected to be independent conditional on the covariate information. Specifically, we compare our LGCP approach estimated using SPDE-INLA with the GAM-based procedure, both with a negative-binomial distribution for the counts (GAM NB) and with the simpler Poisson distribution (GAM Po). As a baseline model we use a homogeneous Poisson model with no covariates, spatial term, or other random effects (Hom Po). 

\subsection{Hooded seals}

Within the flight transect sparsely covering the whelping region, a total of 777 hooded seal pups were counted. The blue dots in Figure \ref{fig:sealdata} show how these are spread on the 2792 photos, with between 0 and 12 pups per photo.

\begin{figure}
	\centering
	\makebox{
		\includegraphics[width=0.49\columnwidth]{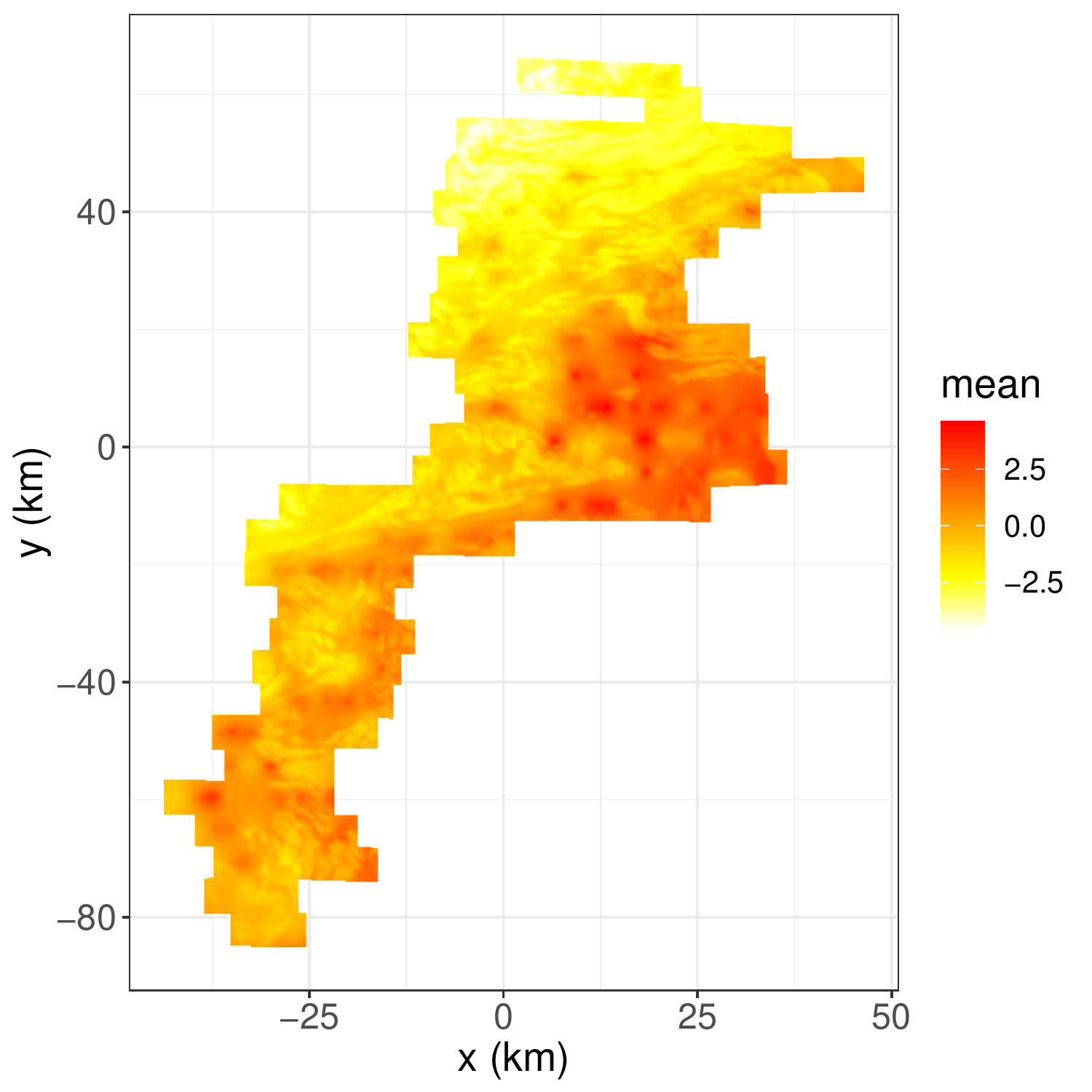}
		\includegraphics[width=0.49\columnwidth]{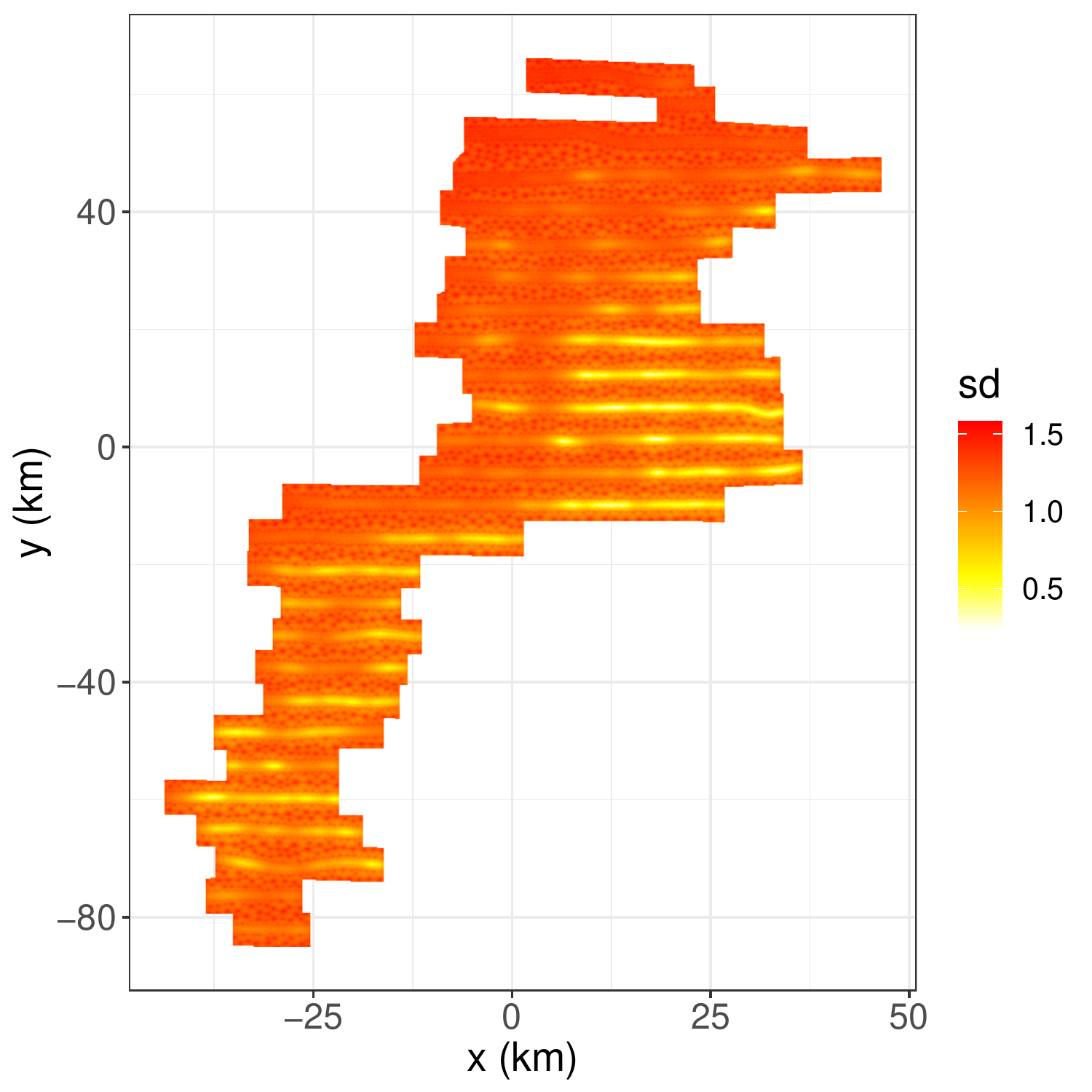}
	}
	\caption{The mean and standard deviation of the random field $Z_f(\bm{s})$ in \eqref{eq:Zf}, fitted for hooded seals with our LGCP approach.}
	\label{fig:latentFieldHooded}
\end{figure}

Figure \ref{fig:latentFieldHooded} shows the mean and standard deviation of the random field fitted using our LGCP procedure outlined in Section \ref{sec:stoch.int.model}. As seen, the latent field captures the high intensity of hooded seal pups in the middle-eastern part of the whelping region. This area has a medium range ice thickness, cf. Figure~\ref{fig:satellitedata}. There is also an increased seal pup intensity further south, in particular closer to the open water. Generally speaking, as seen from the standard deviation plot, the uncertainty is rather large where the intensity is low, while it is smaller where the intensity is high. This means that apart from the north, one is fairly certain that there are \textit{some} seal pups in areas where seal pups are observed nearby, while there could very well exist seal pups in locations where none are observed at the neighboring observation sites.
As a result of very few seals being observed in the north, the mean intensity here is so low that it is unlikely that a significant amount of seals have settled there. 

The \textit{range} of the latent field, defined as the distance at which the spatial correlation is approximately 0.1, has a posterior mean of 3.63 km, or about 2/3 of the distance between two transects. This means that in an area lying between two transects, the latent field is essentially determined by the two neighboring transects. Further, the fitted model gives the following posterior means for the intercept ($\alpha$) and the fixed effects ($\bm{\beta}$): $\text{mean}_{\alpha} = -1.37, \text{mean}_{\bm{\beta},q} = 9.07, \text{mean}_{\bm{\beta},s_1,s_2,s_{12}} = (0.07,-0.05, -0.02)$. 
Similarly the GAM NB approach has the following coefficient estimates: 
$\widehat{\alpha} = -2.58, \widehat{\bm{\beta}}_q = 9.59$, and 
$\widehat{\bm{\beta}}_{s_1, s_2,s_{12}} = (0.06,-0.02, 0.03)$.

\begin{figure}
	\centering
	\makebox{\includegraphics[width=\columnwidth]{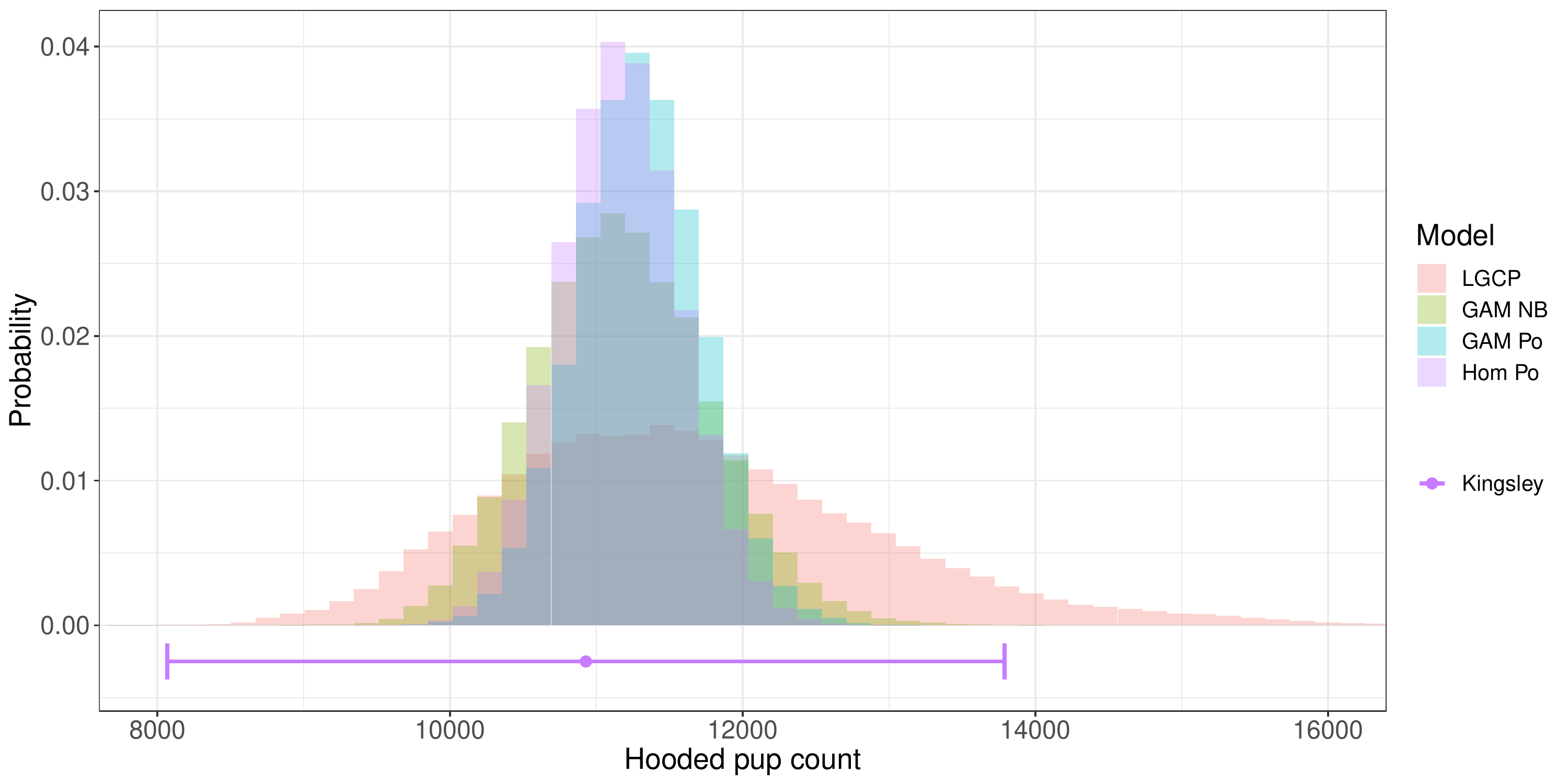}}
	\caption{The posterior predictive distributions for the total hooded pup counts in the whelping region for the five competing models. For Kingsley's method, we show the point estimate +/- 2 times the estimated standard deviation, corresponding to an approximate 95\% confidence interval under a normal distribution assumption.}
	\label{fig:postPredHooded}
\end{figure}

\begin{table}
	\caption{\label{tab:summary_hooded}Summary table for the posterior predictive distributions of the total count of hooded seal pups in the whelping region.} 
	\makebox{\begin{tabular}{lcccccc}
			\hline
			& mean & median & mode & IQR & 0.025-quantile & 0.975-quantile \\ 
			\hline
			LGCP & 11649 & 11503 & 11472 & 1699 & 9472 & 14741 \\ 
			GAM NB & 11178 & 11157 & 11093 & 807 & 10075 & 12395 \\ 
			GAM Po & 11296 & 11292 & 11093 & 572 & 10467 & 12147 \\ 
			Hom Po & 11166 & 11161 & 11172 & 555 & 10372 & 11987 \\ 
			\hline
	\end{tabular}}
\end{table}

Figure \ref{fig:postPredHooded} shows the posterior predictive distribution for the total pup count in the whelping region using our LGCP procedure, along with the corresponding results for the two GAM-based procedures (negative binomial and Poisson response) and the homogeneous Poisson model. A simple summary of Kingsley's method is also given for reference. Table \ref{tab:summary_hooded} summarizes the predictive distributions. While most of the mass coincides for the four different approaches, the GAM NB method yields the lowest estimate and our LGCP method the highest estimate. The Hom Po approach and the GAM Po methods have the lowest prediction uncertainty. Here, our LGCP method has an interquartile range approximately twice as large as the GAM NB approach, and three times larger than GAM Po and Hom Po approaches. For all the methods, the mean predictor is slightly higher than the median predictor, indicating a small degree of skewness with a heavier upper tail. This effect is most pronounced for the LGCP approach.

\begin{table}
	\caption{\label{tab:Validation_hooded}Validation results on photo level (one prediction per photo) and transect level (one prediction per transect), respectively. Lower and upper bounds of 90\% bootstrapped confidence intervals for the scores are shown in parenthesis. Cells shown in italics are the best (smallest) per column. Those which are significantly smaller than the others (defined as having non-overlapping confidence intervals) are also in bold.}
	\centering
	\makebox{\begin{tabular}{lcccc}
			\\
			\multicolumn{5}{c}{\textbf{HOODED SEALS: PHOTO LEVEL}}\\
			\hline
			\multicolumn{1}{c}{} &	\multicolumn{2}{c}{Random 10-fold CV} &	\multicolumn{2}{c}{Leave-out full transect}\\ 
			& CRPS & logScore & CRPS & logScore \\ 
			\hline
			LGCP & \textit{\textbf{0.18 (0.16, 0.19)}} & \textit{0.47 (0.44, 0.49)} & \textit{0.22 (0.20, 0.25)} & 0.54 (0.51, 0.57) \\ 
			GAM NB & 0.21 (0.19, 0.23) & 0.51 (0.47, 0.53) & \textit{0.22 (0.20, 0.24)} & \textit{0.53 (0.50, 0.56)} \\ 
			GAM Po & 0.22 (0.20, 0.24) & 0.54 (0.51, 0.58) & 0.24 (0.22, 0.26) & 0.58 (0.54, 0.62) \\ 
			Hom Po & 0.26 (0.24, 0.28) & 0.77 (0.71, 0.84) & 0.26 (0.24, 0.29) & 0.78 (0.72, 0.85) \\ 
			\hline
			&&&&\\
			\multicolumn{5}{c}{\textbf{HOODED SEALS: AGGREGATE/TRANSECT LEVEL}}\\
			\hline
			\multicolumn{1}{c}{} &	\multicolumn{2}{c}{Random 10-fold CV} &	\multicolumn{2}{c}{Leave-out full transect}\\ 
			& CRPS & logScore & CRPS & logScore \\ 
			\hline
			LGCP& 5.43 (4.04, 6.99) & 3.68 (3.51, 3.86) &  9.91 (5.99, 14.80) & \textit{3.67 (3.26, 4.09)} \\ 
			GAM NB & 5.93 (4.95, 7.00) & 3.79 (3.68, 3.91) &  \textit{9.37 (5.66, 13.63)} & 3.68 (3.11, 4.27) \\ 
			GAM Po & 5.90 (4.49, 7.42) & 3.72 (3.50, 3.96) & 10.14 (5.86, 15.09) & 4.14 (3.32, 5.01) \\ 
			Hom Po & \textit{4.89 (2.21, 10.06)} & \textit{3.58 (3.14, 4.65)} & 20.70 (1.29, 55.68) & 12.81 (2.46, 37.28) \\ 
			\hline
	\end{tabular}}
\end{table}

Table~\ref{tab:Validation_hooded} shows the results from the validation scheme applied to the four methods we compare here, as outlined in Section \ref{sec:verification}. At the photo level, we issue a prediction for the pup count per photo, for either 10\% of the photos or all photos in a single transect at a time. Here, our LGCP method yields very good results, in particular for the random 10-fold cross-validation where observations are generally available in the neighborhood of the prediction locations. It is significantly the best method for this setting under the CRPS, defined as having non-overlapping 90\% CI strictly below all others, and almost significant in terms of the logScore. When leaving out a full transect at a time, our LGCP method and GAM NB method perform very similar, and somewhat better than the others.

At the aggregate/transect level we issue a joint prediction for the total pup count per 10\% of the photos or per transect, respectively. Here, the baseline homogeneous Poisson model performs very well with the random leave-out scheme, indicating that the data may be close to homogeneous across the photos.  However, this does not transfer to the second set-up where we leave out one transect at a time in which case our LGCP method and the GAM NB method again perform well. As seen in Figure~\ref{fig:sealdata}, the data do not appear homogeneous across the transects. Note that these reported scores are averages over relatively few predictions, only 10 distinct predictions for the random 10-fold cross-validation study and 27 distinct predictions for the leave-out-transect setup. The score values are thus associated with a large degree of uncertainty. The uncertainty is particularly high for the homogeneous Poisson model that, due to the assumption of homogeneity, is especially sensitive to inhomogeneities across the different cross-validation sets.  

\begin{figure}
	\centering
	\makebox{\includegraphics[width=\columnwidth]{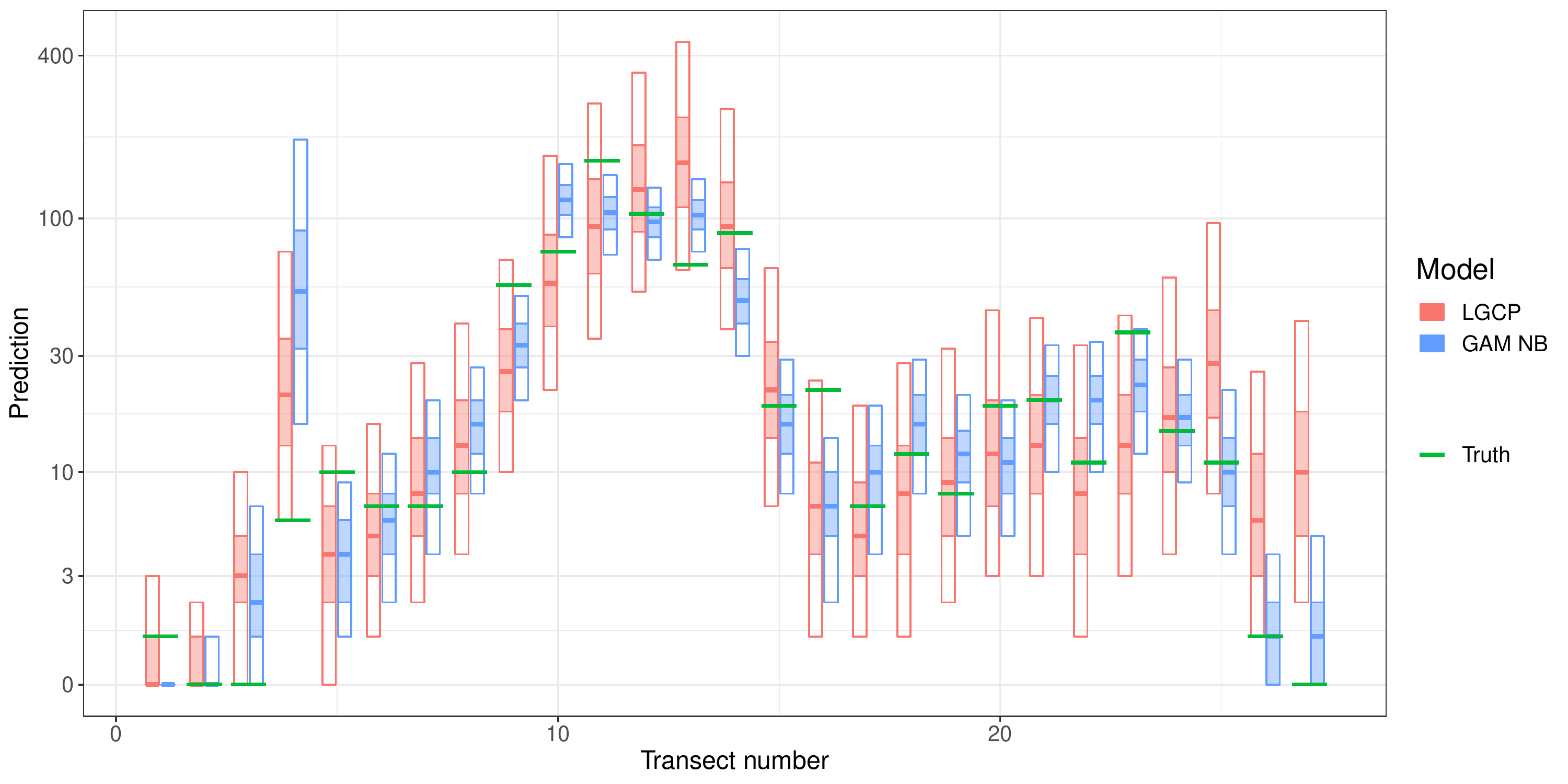}}
	\caption{The posterior predictive distributions per transect for the LGCP and GAM NB methods plotted against the true count for hooded seals. For each method, the solid line shows the median, the light colored box shows the 50\% CI, while the transparent box shows the 90\% CI. The y-axis has a $\log_{10}(x+2)$-scale to better show differences.}
	\label{fig:transectResHooded}
\end{figure}

Based on the results in Table~\ref{tab:Validation_hooded}, our LGCP method performs somewhat similar to the GAM NB method and these are both clearly superior to the other two alternatives; the two methods can only be distinguished in terms of CRPS on the photo level. Despite this, their resulting posterior predictive distributions are quite different. To get further insight into this phenomenon, Figure~\ref{fig:transectResHooded} shows the posterior predictive distributions for the two methods per transect in the leave-out-transect setup, plotted against the true transect counts. As seen from the figure, the GAM NB method seems to have too narrow credibility intervals, while the LGCP approach appears more calibrated. In fact, the 90\% CI covers the true count in $26/27\approx 96\%$ of the transects for the LGCP approach, and only $18/27\approx 67\%$ of the transects for the GAM NB approach. The 50\% CI is covered in respectively $16/27\approx 59\%$ and $11/27\approx 41\%$ of the transects for the LGCP and GAM NB approaches. Note that when looking at the MAE (mean absolute error) of the median count estimate for the two methods, the GAM NB method achieves MAE of 12.4, while the LGCP method is slightly worse with MAE of 13.8. Thus, the GAM NB method seems to do somewhat better as a point estimator, while the LGCP is better at estimating the uncertainty.

\subsection{Harp seals}

A total of 6034 harp seal pups where observed on the 2792 photos from the aerial photographic survey. As illustrated by the red dots in Figure \ref{fig:sealdata}, there are much larger packs of harp seal pups than hooded seal pups, indicating a higher degree of inhomogeneity. Here, the pup count per photo ranges from 0 to 160. 

\begin{figure}
	\centering
	\makebox{
		\includegraphics[width=0.49\columnwidth]{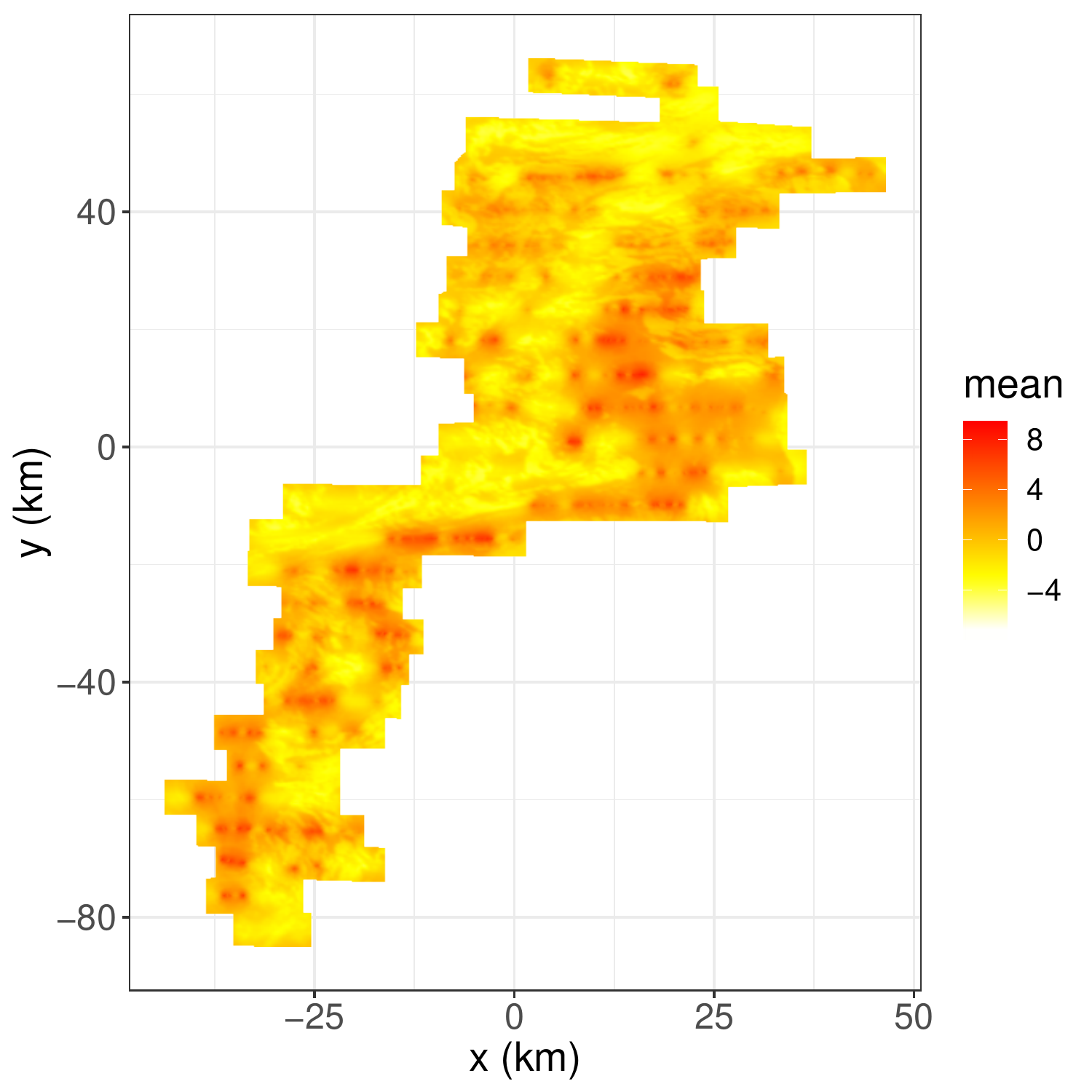}
		\includegraphics[width=0.49\columnwidth]{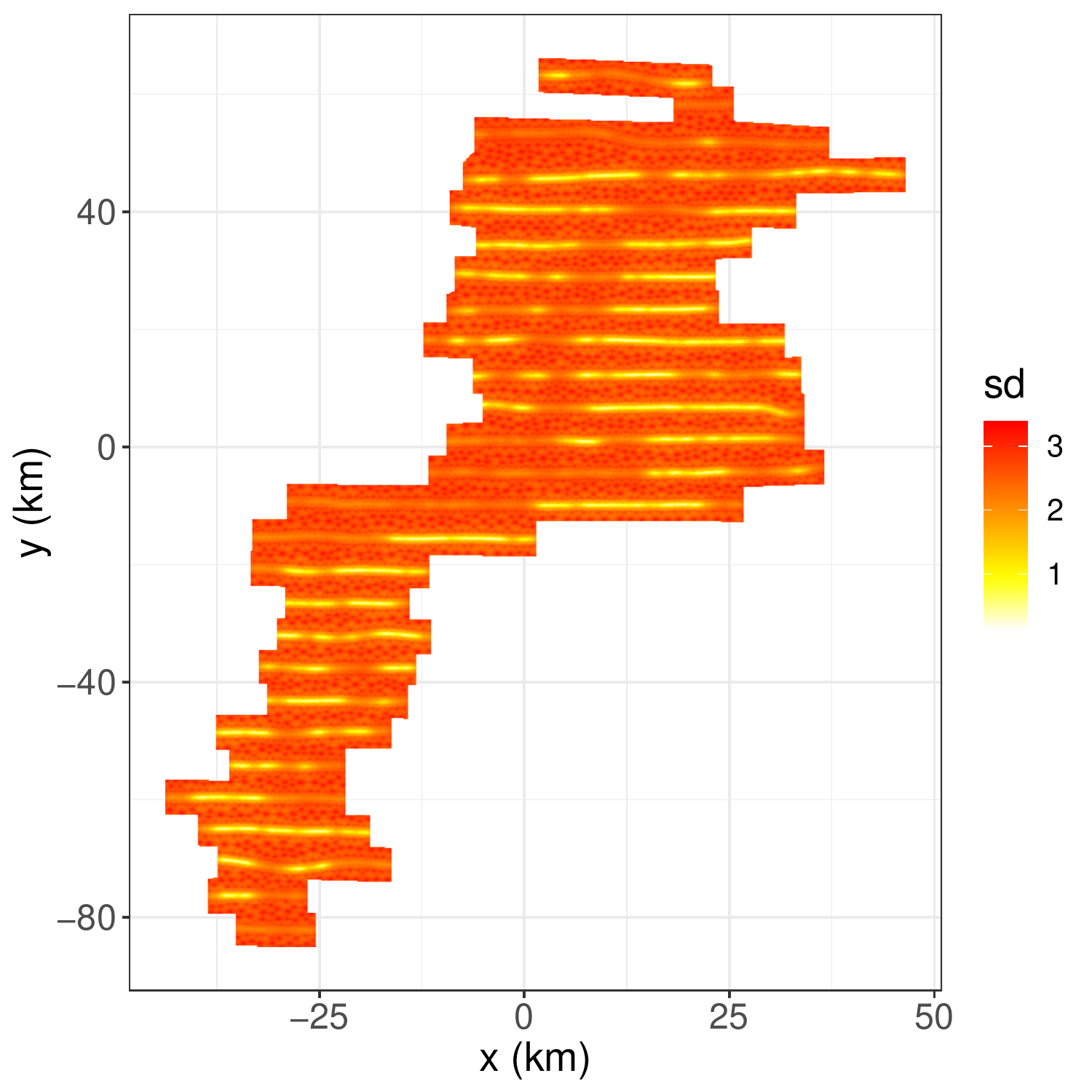}
	}
	\caption{The mean and standard deviation of the random field $Z_f(\bm{s})$ in \eqref{eq:NiZ} fitted for harp seals with our LGCP approach.}
	\label{fig:latentFieldHarps}
\end{figure}

Figure \ref{fig:latentFieldHarps} shows the mean and standard deviation of the fitted random field $Z_f(\bm{s})$ using our LGCP procedure. Compared to the latent field for the hooded seal pups in Figure \ref{fig:latentFieldHooded}, the mean field here has a much higher degree of spatial variation with higher and steeper peaks. However, the locations where the seal pups mainly appear are similar to those for the hooded seal pups, except for some additional colonies in the north and north-west of the region. Otherwise, the properties of the two fields are fairly similar. The range of the latent field has a posterior mean of 2.89 km, a slightly smaller value than for the hooded seals which corresponds to roughly half the distance between two transects. The fitted model gives the following posterior means for intercept ($\alpha$) and fixed effects ($\bm{\beta}$): $\text{mean}_{\alpha} = -2.77, \text{mean}_{\bm{\beta},q} =14.70, \text{mean}_{\bm{\beta},s_1,s_2,s_{12}} = (0.03, 0.01, -0.003)$. Note that the covariate effects are stronger for the harps than the hooded seals. This is natural as there are many more observed harps than hooded seals.
The GAM NB approach gives the following coefficient estimates: 
$\widehat{\alpha} = 2.70, \widehat{\bm{\beta}}_q = 19.05$, and $\widehat{\bm{\beta}}_{s_1, s_2,s_{12}} = (0.02, 0.04, -0.1)$.

\begin{figure}
	\centering
	\makebox{\includegraphics[width=\columnwidth]{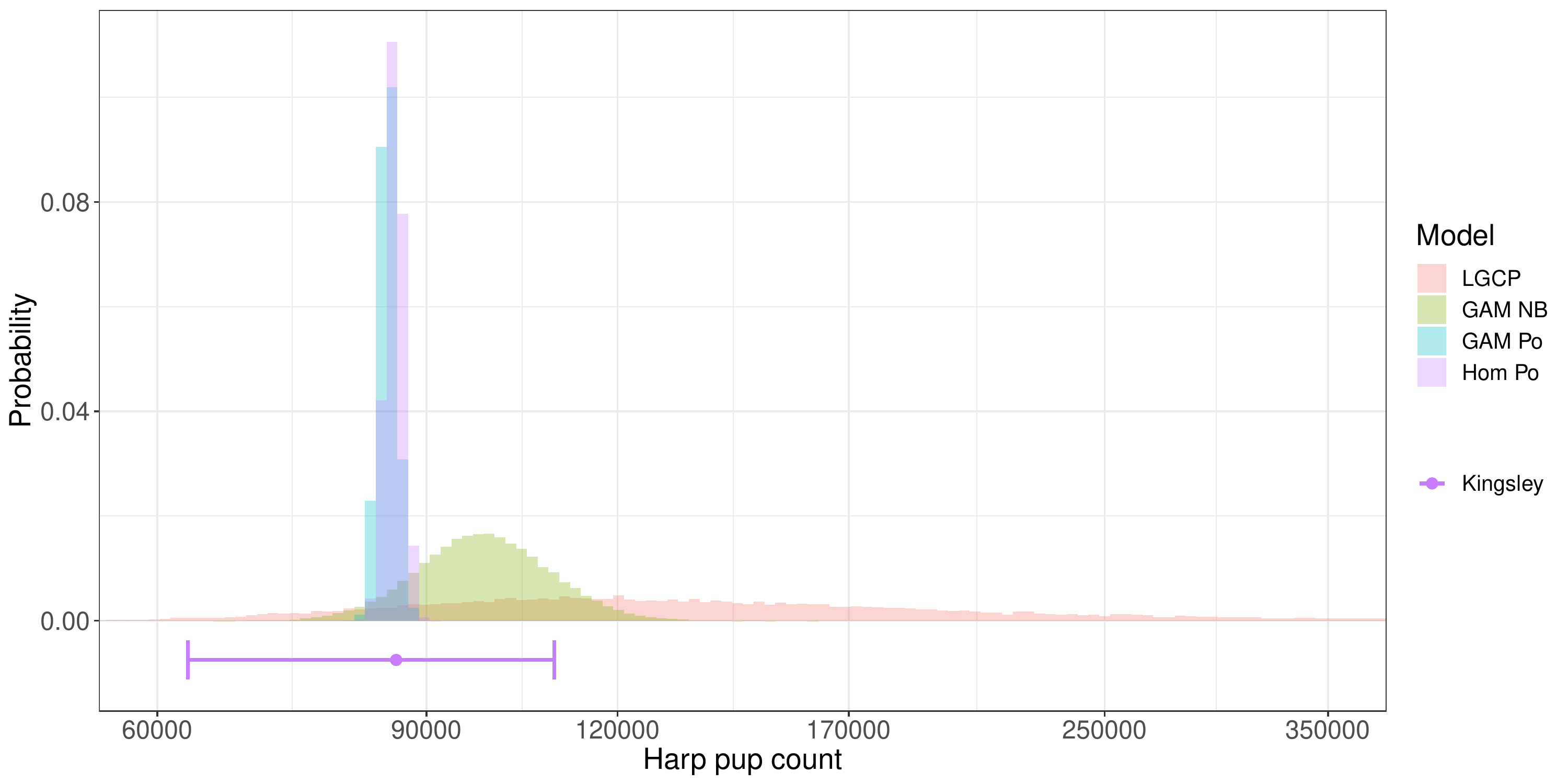}}
	\caption{The posterior predictive distributions for the total harp pup counts in the whelping region for the five different approaches. The x-axis is plotted on log-scale. For Kingsley's method, we show the point estimate +/- 2 times the estimated standard deviation, corresponding to an approximate 95\% confidence interval under a normal distribution assumption.}
	\label{fig:postPredHarps}
\end{figure}

Figure \ref{fig:postPredHarps} shows the posterior predictive distribution for the total number of harp seal pups in the whelping region using our LGCP procedure, along with the corresponding results for the two GAM procedures (GAM NB and GAM Po) and the homogeneous Poisson model (Hom Po). A simple summary of Kingsley's method is also given for reference. Table \ref{tab:summary_Harps} summarizes these distributions. The LGCP prediction of the total harp seal pup count is highly uncertain, essentially saying that the total number of seal pups could very well be above 250 000, but also less than 100 000. In contrast, the GAM NB method's upper tail ends at about 130 000 seal pups, while the GAM Po and Hom Po methods agree that there are between 80 000 and 90 000 harp seal pups within the whelping region. Thus, there are significant differences both between the centrality and width of the different methods' posterior predictive distributions. The predictive distributions for GAM Po and Hom Po are essentially symmetric while the one for GAM NB is minimally skewed with a heavier upper tail. The LGCP approach, however, yields a severely skewed predictive distribution with a predictive mean that is 15\% larger than the predictive median due to the large uncertainty in the estimation of the random field $Z_f$, cf. Figure~\ref{fig:latentFieldHarps}. 

\begin{figure}[ht!]
	\centering
	\includegraphics[width=0.95\columnwidth]{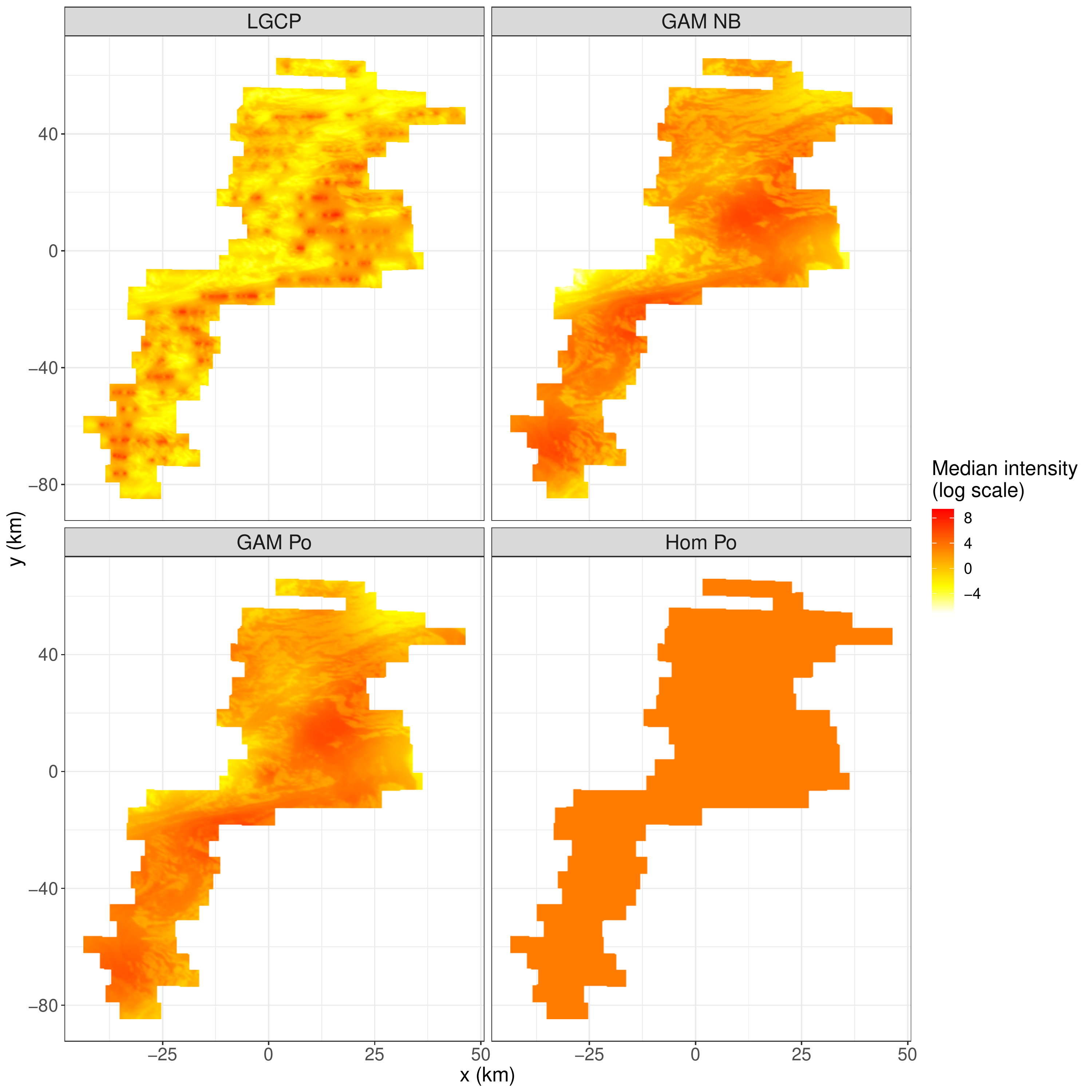}
	\caption{The estimated posterior median intensity field (plotted on a logarithmic scale) for all four methods fitted for harp seals.}
	\label{fig:mean_intensity_log_scale_harps_all_methods}
\end{figure}

\begin{table}
	\caption{\label{tab:summary_Harps}Summary table for the posterior predictive distributions for the total number of harp seal pups in the whelping region.} 
	\makebox{\begin{tabular}{lcccccc}
			\hline
			& mean & median & mode & IQR & 0.025-quantile & 0.975-quantile \\ 
			\hline
			LGCP & 147919 & 127965 & 110996 & 72347 & 69267 & 357185 \\ 
			GAM NB & 98617 & 98035 & 91876 & 12895 & 81023 & 119349 \\ 
			GAM Po & 84852 & 84852 & 84910 & 1536 & 82681 & 87094 \\ 
			Hom Po & 85751 & 85747 & 85719 & 1539 & 83529 & 88004 \\ 
			\hline
	\end{tabular}}
\end{table}

To further investigate the differences between the four predictions, we have plotted the posterior median intensity fields for the four methods in Figure \ref{fig:mean_intensity_log_scale_harps_all_methods}, cf. the expressions for the intensity fields in the third column in Table \ref{tab:models}. The regression coefficient for the ice density covariate $q$ shown in Figure~\ref{fig:satellitedata} is somewhat higher for the GAM approaches than for the LGCP approach (posterior mean of 19.05 for GAM NB compared to 14.70 for LGCP), resulting in fairly smooth median intensity fields that reflect the spatial structure of the ice density field. The median LGCP intensity field, however, has stronger inhomogeneities across space and appears more strongly influenced by the data density shown in Figure~\ref{fig:sealdata} with noticeable peaks in locations with higher observed seal pup density. Overall, however, the median intensity is higher for the GAM methods than for the LGCP. The reason that LGCP still produces higher predictions (cf.~Table \ref{tab:summary_Harps}) is partly due to the high peaks, but mainly due to the high degree of uncertainty. 
The posterior of $Z(\bm{s})$ is more or less symmetric with a high degree of uncertainty, such that large values are sampled quite frequently. These large values boosts the intensity $\lambda(\bm{s})=\exp(Z(\bm{s}))$ considerably, resulting in large predictions. The lower uncertainty of the GAM methods does not have the same effect. Further, note that although the two GAM models have very similar posterior median intensities, their posterior predictive distributions are quite different, cf. Figure \ref{fig:postPredHarps} and Table \ref{tab:summary_Harps}. 

\begin{figure}
	\centering
	\makebox{\includegraphics[width=\columnwidth]{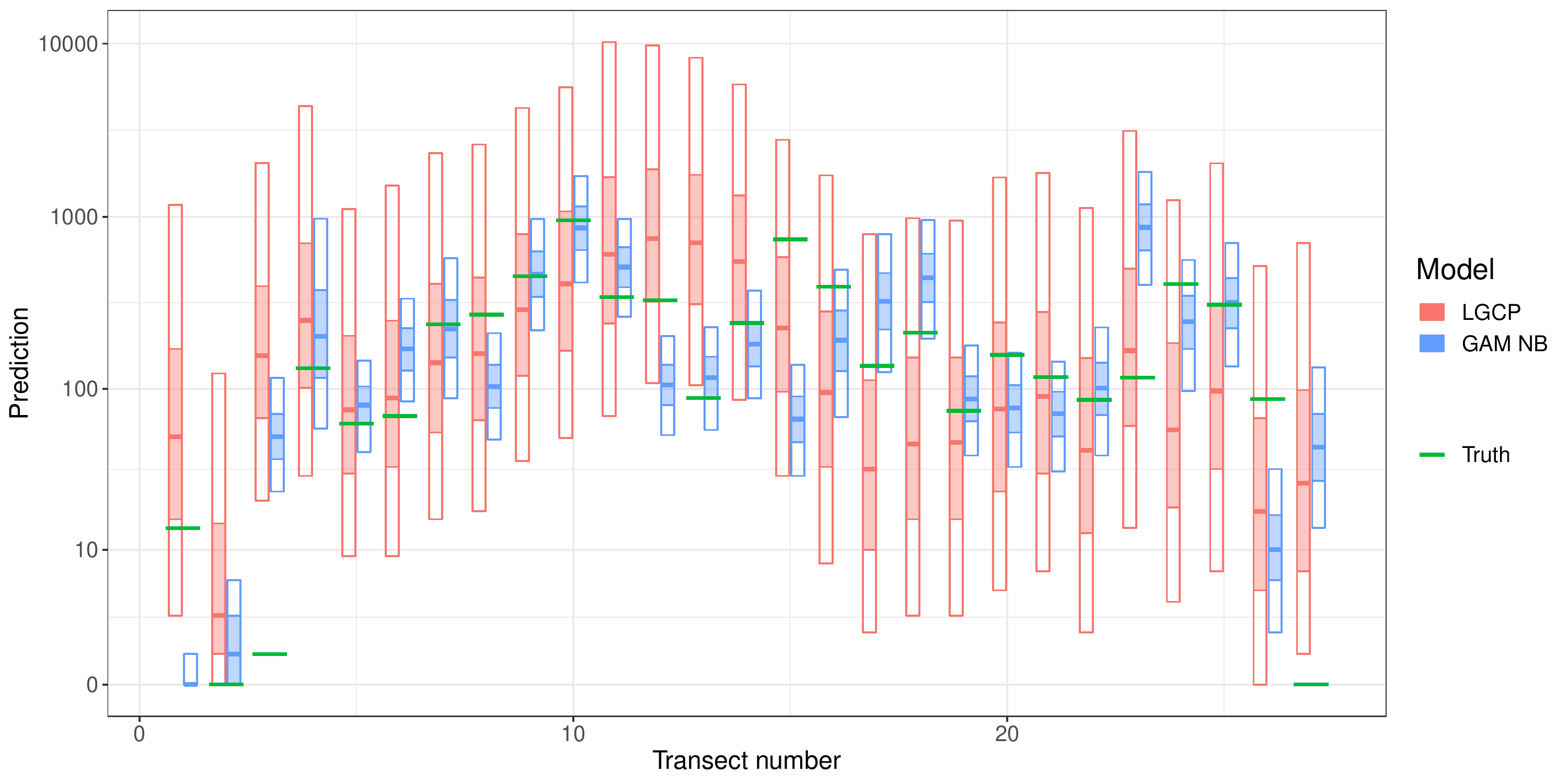}}
	\caption{The posterior predictive distributions per transect for the LGCP and GAM NB methods plotted against the true count for harp seal pups. For each method, the solid line shows the median, the light colored box shows the 50\% CI, while the transparent box shows the 90\% CI. The y-axis takes a $\log_{10}(x+2)$-scale to better show differences.}
	\label{fig:transectResHarps}
\end{figure}

As for the hooded seals, we take a closer look at the LGCP and GAM NB methods to better understand how well their different estimates of the prediction uncertainty match the actual uncertainty, see Figure~\ref{fig:transectResHarps}. As expected, the GAM NB method has a much narrower credibility intervals which too often fail to cover the true seal pup count in the transect, while our LGCP method shows much better calibration.  Out of the 27 transects, the 90\% credibility intervals for the LGCP and GAM NB approaches covers the true count in respectively $24/27 \approx 89\%$ and $18/27 \approx 67\%$ of the transects. The corresponding coverage for the 50\% interval are $14/27 \approx 52\%$ and $11/27 \approx 41\%$, respectively. Thus, it is clear that the GAM NB method is underdispersive, while our procedure seems well calibrated. On the other hand, as for the hooded seals, the posterior median of the GAM NB method does better as a point estimator in terms of MAE (130.2) than our LGCP method (179.1).

Table \ref{tab:Validation_harps} shows the results from the validation procedure for the harp seals, yielding similar model rankings as for the hooded seals. On photo level, the LGCP approach gives a significantly better CRPS and logScore under random 10-fold cross-validation. Leaving out full transects gives no significantly best method although the GAM Po method tends to generally do well here. However, as before, these average scores are associated with a very high degree of uncertainty so that a ranking of the methods based on these results is not advisable. 

\begin{center}
	
	\begin{table}
		\caption{\label{tab:Validation_harps}Validation results on photo level (one prediction per photo) and transect level (one prediction per transect), respectively. Lower and upper bounds of 90\% bootstrapped confidence intervals for the scores are shown in parenthesis. Cells shown in italics are the best (smallest) per column. Those which are significantly smaller than the others (defined as having non-overlapping confidence intervals) are also in bold.}
		\centering
		\begin{tabular}{lcccc}
			\\
			\multicolumn{5}{c}{\textbf{HARP SEALS: PHOTO LEVEL}}\\
			\hline
			\multicolumn{1}{c}{} &	\multicolumn{2}{c}{Random 10-fold CV} &	\multicolumn{2}{c}{Leave-out full transect}\\ 
			& CRPS & logScore & CRPS & logScore \\ 
			\hline
			LGCP & \textbf{\textit{1.14 (1.01, 1.27)}} & \textbf{\textit{0.95 (0.91, 1.00)}} & 1.96 (1.72, 2.20) & 1.28 (1.22, 1.33) \\ 
			GAM NB & 1.78 (1.58, 2.00) & 1.17 (1.11, 1.22) & \textit{1.90 (1.67, 2.13)} & \textit{1.27 (1.21, 1.33)} \\ 
			GAM Po & 2.32 (2.10, 2.55) & 2.09 (2.00, 2.17) & 2.46 (2.22, 2.71) & 2.17 (2.08, 2.26) \\ 
			Hom Po & 2.64 (2.40, 2.90) & 3.47 (3.28, 3.67) & 2.66 (2.42, 2.92) & 3.49 (3.30, 3.69) \\ 
			\hline &&&&\\
			\multicolumn{5}{c}{\textbf{HARP SEALS: AGGREGATE/TRANSECT LEVEL}}\\
			\hline
			\multicolumn{1}{c}{} &	\multicolumn{2}{c}{Random 10-fold CV} &	\multicolumn{2}{c}{Leave-out full transect}\\ 
			& CRPS & logScore & CRPS & logScore \\ 
			\hline
			LGCP & 95.98 (51.20, 148.78) & 6.57 (5.93, 7.33) & 152.95 (111.93, 198.59) & \textit{6.49 (5.98, 6.94)} \\ 
			GAM NB & 88.33 (70.94, 106.87) & \textit{6.45 (6.31, 6.62)} &  96.70 (61.05, 139.91) & 7.00 (6.24, 7.76) \\ 
			GAM Po & \textit{60.93 (42.57,  79.87)} & 8.03 (6.83, 9.22) &  \textit{55.96 (39.62,  74.51)} & 8.42 (7.34, 9.39) \\ 
			Hom Po & 102.62 (8.87, 296.09) & 62.92 (4.31, 300.60) &  149.38 (7.70, 454.53) & 82.63 (4.08, 251.00) \\ 
			\hline
		\end{tabular}
	\end{table}
\end{center}

\section{Conclusions and discussion}
\label{sec:ConcludingRemarks}

We have presented a point process based approach to estimate seal pup production based on observational data from an aerial photographic survey. Using the SPDE-INLA framework, we fit a Bayesian hierarchical model with Poisson counts following a log-Gaussian Cox process (LGCP) model formulation. As an additional contribution to seal pup production estimation, we adopt the use of satellite imagery as covariates in the modeling process, to act as a proxy for ice thickness. The approach is applied to 2012 survey data from the Greenland Sea, with both harp and hooded seal pup counts, and compared to several reference methods that can be associated with non-homogeneous or homogeneous point process formulations rather than the doubly stochastic setting of the Cox process. 

The competing methods are compared in two cross-validation studies. The proposed LGCP approach generally performs best locally, while no method stands out as the best on a more regional scale. However, this lack of discrimination in the comparison at the regional scale is not surprising given the relatively small size of our data set resulting in large uncertainties in the scores, see e.g. the discussion and examples in \citet{thorarinsdottir2019verification}.  The most distinguishing characters of the LGCP  method are higher count predictions and a large prediction uncertainty compared to the other methods. Our analysis suggests that the wide uncertainty bounds are indeed necessary to issue calibrated predictions. This further suggests that the amount of data collected in the aerial photographic survey might not be sufficient to obtain reliable estimates of the total seal pup production in the whelping region. 

The results show that the model uncertainty directly affects the seal pup count predictions to a large degree. This exemplifies the importance of a proper modeling of all involved uncertainties. In particular, a comparison of the two GAM approaches for the harp seals shows that even if the posterior median intensity fields for the two approaches are very similar, the resulting posterior predictive distributions for the total seal pup counts are quite different. This indicates that it is not sufficient to consider the mean behavior. \citet{salberg2009estimation} found that a negative binomial likelihood was necessary in order to obtain a good fit to the data. However, the conclusions of our analysis indicate that a more careful and comprehensive assessment of both the model uncertainty and the spatial inhomogeneities may warrant the use of the simpler Poisson likelihood, in particular if the doubly stochastic framework of the LGCP is applied.     

A direct comparison of the models based on a non-homogeneous versus a doubly stochastic point process formulation is somewhat complicated by the fact that both the models and the estimation approaches differ. In particular, the SPDE-INLA approach is an approximate fully Bayesian approach with priors on all parameters, while the GAM approach is merely a pseudo-Bayesian procedure. In particular, we do not include smoothing parameter uncertainty in the GAM approach. While the GAM approach has been extended in this direction, see e.g. \citet{wood2016just}, such an extension has yet to be implemented for the negative binomial likelihood. We have thus aimed to replicate the implementation of \citet{salberg2009estimation} for the comparison.

Notably, the predictive distributions for the total seal pup counts are very similar under the homogeneous Poisson model estimated with a fully Bayesian approach and under the non-homogeneous Poisson model estimated with a pseudo-Bayesian approach. This invites the conclusion that the vast difference between these two predictive distributions and that under the doubly stochastic Poisson model is, to a large degree, due to the inclusion of the random effect in the intensity function of the doubly stochastic model rather than choice of inference approach. 

For the LGCP inference, we have applied the SPDE-INLA approach directly as implemented in the {\tt R} package {\tt INLA} \citep{lindgren2011explicit,Rue2009}. Alternatively, the more recent package {\tt inlabru} \citep{bachl2019inlabru} is based on the SPDE-INLA software to model point pattern data from surveys with varying detection probabilities over the sampled area. Our setting is slightly different, in that the detection probability is assumed to be equal to one over the entire sampled area and we only have sampled counts per each photo rather than the precise locations of the seal pups. While this somewhat simpler setting can also be analyzed with {\tt inlabru}, we have chosen to implement out own version which allows for a slightly higher flexibility in the generation of the mesh. Besides small differences in the manner in which edge effects are treated when approximating the integral over the latent field, our implementation should give comparable results to {\tt inlabru}.     

In our model specification, we only consider linear effects of the satellite covariate and linear spatial effects. We investigated including non-linear effects and squared terms both for the LGCP and the deterministic intensity model formulations. However, as this did not improve the performance of the models, such terms were not included in the final model specifications. 


In the present work, harp and hooded seals have been modeled separately, based on a single survey. As an alternative, one may consider building a joint model for harp and hooded seals, for instance by using a common spatial field, in addition to seal specific ones (see e.g.~\citet{waagepetersen2016analysis}). Further, due to drifting ice and moving seals, the spatial locations of the seal pups cannot be directly compared from one survey to the next. However, most of the seals tend to stay in more or less the same packs from one year to another. Such information could potentially be utilized to construct informative priors, which may reduce the modeling uncertainty. 
It would be interesting to see investigations on such attempts at borrow strength, either from previous surveys or between seal types.
Such investigations are, however, out of scope of the present paper.

\section*{Acknowledgments}
Martin Jullum and Thordis Thorarinsdottir acknowledge the support of The Research Council of Norway through grant 240838 ``Model selection and model verification for point processes''. Fabian Bachl was supported by EPSRC grant EP/K041053/1 ``Modelling spatial distribution and change from wildlife survey data''. We thank Tor Arne {\O}ig{\aa}rd for helpful discussions and for assisting with the photographic survey data which were provided by the Norwegian Institute of Marine Research, {\O}ivind Due Trier for retrieving the satellite image data, Dawid Lawrence Miller and Alex Lenkoski for helpful discussions regarding the various inference approaches.     

\appendix

\section{The integrated nested Laplace approximation}
\label{app:INLA}
The integrated nested Laplace approximation (INLA) methodology proposed by \citet{Rue2009}, and implemented in the {\tt R}-package {\tt INLA} (\url{www.r-inla.org}), allows for computationally feasible approximate Bayesian inference for latent Gaussian models. In latent Gaussian models, $n$ univariate observations $\bm{y}=(y_1,\ldots,y_n)^\top$ are assumed to be conditionally independent given $m$ latent Gaussian variables $\bm{z}=(z_1,\ldots,z_m)^\top$ and a set of hyperparameters $\bm{\theta}$. More precisely, the INLA implementation covers models of the form 
\begin{align}\label{eq:INLA.model}
p(\bm{y}|\bm{z},\bm{\theta}) &= \prod_{i=1}^n p(y_i|\eta_i,\bm{\theta}), \text{ with }\eta_i = \sum_{j=1}^m c_{ij}z_j \text{ for fixed $c_{ij}$}, \notag \\
p(\bm{z}|\bm{\theta}) &\sim \N(\bm{\mu}(\bm{\theta}),Q(\bm{\theta})^{-1}),  \\
\bm{\theta} &\sim p(\bm{\theta}), \notag
\end{align}
where the latent variables $\bm{z}$ may depend on additional (fixed) covariates, for a large class of models for $\bm{y}$.

For computationally fast inference it is essential that the precision matrix $Q(\bm{\theta})$ is sparse and that the parameter vector $\bm{\theta}$ is of a fairly low dimension. This covers models where the latent field is a {\em Gaussian Markov random field} (GMRF). For the inference, INLA utilizes several nested Laplace approximations. That is, the posterior distribution of $\bm{\theta}$ is approximated by 
\begin{align}
p(\bm{\theta}|\bm{y}) \approx \tilde{p}(\bm{\theta}|\bm{y}) \propto \frac{p(\bm{y},\bm{z},\bm{\theta})}{p_G(\bm{z}|\bm{y},\bm{\theta})}\bigg |_{\bm{z}=\bm{z}^*(\theta)}, \label{eq:post.theta}
\end{align}
where $p_G(\bm{z}|\bm{y},\bm{\theta})$ is a Gaussian approximation to the full conditional distribution of $\bm{z}$, and $\bm{z}^*(\bm{\theta})$ is the mode of $p(\bm{z}|\bm{y},\bm{\theta})$ for a given $\bm{\theta}$. The marginals of this low-dimensional posterior distribution are typically computed by direct numerical integration. 
The marginals for the latent field, $p(z_j|\bm{y})$, are typically computed by first obtaining a Laplace approximation $\tilde{p}(z_j| \bm{\theta}, \bm{y})$ similar to \eqref{eq:post.theta}, or a Taylor approximation of that distribution, and then solve $\int \tilde{p}(z_j| \bm{\theta}, \bm{y})\tilde{p}(\bm{\theta}|\bm{y})\dd \bm{\theta}$ by numerical integration. See \citet{Rue2009} and \cite{martins2013bayesian} for further details.

\bibliographystyle{biometrika_new}
\bibliography{bibliography}

\end{document}